\documentclass{ws-ijmpa}
\usepackage[super,compress]{cite}
\usepackage{graphicx}
\begin{document}
\markboth{Z.-H. Weng}{Some properties of dark matter field}

%
\catchline{}{}{}{}{}
%

\title{Some properties of dark matter field in the complex octonion space
}

\author{Zi-Hua Weng
}

\address{School of Physics and Mechanical \& Electrical Engineering, Xiamen University, \\ Xiamen, 361005, China
\\
xmuwzh@xmu.edu.cn}

%

\maketitle

\begin{history}
\received{Day Month Year}
\revised{Day Month Year}
\end{history}

\begin{abstract}
The paper aims to consider the electromagnetic adjoint-field in the complex octonion space as the dark matter field, describing some properties of dark matter, especially the origin, particle category, existence region, and force and so forth. Since J. C. Maxwell applied the algebra of quaternions to depict the electromagnetic theory, some scholars adopt the complex quaternion and octonion to study the physics property of electromagnetic and gravitational fields. In the paper, by means of the octonion operator, it is found that the gravitational field accompanies with one adjoint-field, which property is partly similar to that of electromagnetic field. And the electromagnetic field accompanies with another adjoint-field, which feature is partly similar to that of gravitational field. As a result the electromagnetic adjoint-field is able to be chosen as one candidate of the dark matter field. According to the electromagnetic adjoint-field, it is able to predict a few properties of dark matter, for instance, the particle category, interaction intensity, interaction distance, and existence region and so forth. The study reveals that the dark matter particle and gravitational resource both will be influenced by the gravitational strength and force. The dark matter field is capable of making a contribution to physics quantities of gravitational field, including the angular momentum, torque, energy, and force and so on. Further there may be comparatively more chances to discover the dark matter in some regions with the ultrastrong field strength, surrounding the neutral star, white dwarf, galactic nucleus, black hole, and astrophysical jet and so on.

\keywords{dark matter; force; octonion space; gravitational field; electromagnetic field.}
\end{abstract}

\ccode{PACS numbers: 95.35.+d; 11.10.Kk; 02.10.De; 04.50.Kd; 98.62.Dm}


\section{Introduction}	

The concept of dark matter is attracting the attention of many people. Some scholars are fascinated by the dark matter, but bewildered by the concept at the same time. We do not know what the dark matter is, how it is generated, where it exists in, and whether it is one particle. There is neither the effective theory with practical guidance, nor the systematic observational and experimental methods yet. Until recently the appearance of the electromagnetic and gravitational theories describing with the complex octonion replies to some of these puzzles. According to this field theory, a few features of dark matter are able to be predicted, including the origin, particle category, existence region, and force and so forth.

In 1932, Jan Oort postulated it was possible that one `non-visible form of matter' might exist in the galaxy \cite{kuijken} , accounting for the orbital velocities of stars in the Milky Way. In 1933, Fritz Zwicky inferred that there might be one `missing mass' in the galaxy cluster \cite{zwicky} , by means of the observation on the velocities of galaxies in clusters. In 1939, in his Ph.D. thesis, Horace W. Babcock reported the measurements of the rotation curve for the Andromeda Nebula.

In the late 1960s and early 1970s, Vera Rubin measured a lot of galactic rotation curves, attributing the anomalous observations to the existence of dark matter. The observations and calculations of Vera Rubin and Kent Ford showed that most galaxies contain about six times as much mass of `dark matter' as that of ordinary matter \cite{rubin1} . Subsequently many observations of other astronomers corroborate the presence of dark matter in the universe \cite{kusenko} . By the 1980s, the observation of galactic rotation curves has been collecting over the decades, providing the robust observational evidence for the existence of dark matter \cite{bertone} . As a result, most astrophysicists accepted the viewpoint of the existence of dark matter eventually \cite{gupta} .

\subsection{Dark matter detection}

In recent years, the astronomers make a survey of dark matter on a large scale, achieving a number of observations. Meanwhile the alternative approaches are increasing continuously. At present, the observation approach to the detection of dark matter consists of the gravitational lensing \cite{dietrich} , the galactic rotation curve of the spiral galaxy \cite{rubin2} , the velocity dispersions of galaxies in the galaxy cluster (or stars in the galaxy) \cite{faber} , the X-ray hot gas hydrostatic \cite{koopmans} in the galaxy or cluster, the Sunyaev-Zel'dovich effect of the cluster \cite{diego} and so forth. Those observation approaches make numerous and conclusive measurements, resulting in more and more scholars to accept the view point related to the existence of dark matter \cite{dekel} .

The observational achievements in the astronomy empower the scholars to undertake some correlative investigations in the laboratory, attempting to discover dark matter on the microscopic scale. Presently the experiments to detect dark matter in the lab can be divided into two classes: direct detection experiments, and indirect detection experiments. (1) Direct detection experiment. The majority of present experiments mainly use one of two detectors: cryogenic detectors, and Noble liquid detectors. Some with the cryogenic detector embody CDMS \cite{kozaczuk} , CRESST \cite{cresst} , EDEDWEISS \cite{edelweiss} , and EURECA \cite{munster} and so on. Some others with the cryogenic detector contain ZEPLIN \cite{ghag} , XENON \cite{aprile} , DEAP \cite{amaudruz} , ArDM \cite{badertscher} , WARP \cite{weinberg} , DarkSide \cite{agnes} , LUX \cite{lux} , and PandaX \cite{pandax} and so forth. Other detection experiments include SIMPLE \cite{felizardo} and PICASSO \cite{archambault} and so on. (2) Indirect detection experiment. The most experiments search for the products (the gamma rays and particle-antiparticle pairs and so on) of dark matter annihilation or decay. They cover PAMELA \cite{adriani} and AMS \cite{ams} on the International Space Station and so on.

However there is a big difference between the observation and experiment. By contrast with the achievement that there are a lot of astronomic observations, the experiments in the laboratory have not made compelling and robust measurements up to now. Similarly there is not the physics or astronomic theory capable of practical guidance to the detection experiments deal with dark matter either. We are still only at the foothills of pursuing the effect theory of dark matter.

\subsection{Dark matter theory}

For the observations of galaxies and clusters, some alternative theories proposed to account for `missing mass' can be divided into two classes: modified theories, and dark matter theories.

(1) Modified theory. These theories claim that it is possible to modify or extend the existing theories, explaining the galaxy observations without the need for a large amount of undetected matter. These main theories are as follows. (i) Modified Newtonian dynamics. Most of these theories deem that the gravitational theories established by I. Newton and A. Einstein are not perfect enough, so it is necessary to modify the laws of gravity \cite{angus} , solving the problem about the galactic rotation curves and so forth. (ii) Quantum gravity. These theories attempt to unify the gravitational theory and the quantum theory and so on, including the Loop Quantum Gravity \cite{sahu} , Superstring theory \cite{safarzadeh} , and M-theory \cite{rama} and so on. (iii) Quantum vacuum. In the quantum vacuum \cite{dellieu} , under certain conditions, the virtual gravitational dipoles can produce the extra gravity attributing to the dark matter usually. (iv) Topological defeats. The topology of quantum fields may possess some primordial defects, containing the energy and therefore additionally gravitating towards the galaxy \cite{derevianko} . (v) Mass in extra dimension. In certain multidimensional theories \cite{landsberg} , the gravitational force is the unique force capable of effects across all the various extra dimensions, but the forces of other fundamental fields (electromagnetism, strong interaction, and weak interaction) would not be able to cross into extra dimensions. (vi) Quantised inertia. The theory reckons that the inertia is assumed to be due to the effect of horizons on Unruh radiation, resulting in the galaxy rotation \cite{mcculloch} without dark matter. Unfortunately these categories of modified theories did not successfully explain the observations of galaxy cluster up to now.

(2) Dark matter theory. These theories believe that there may be the undetected matter at the moment, including the dark matter and dark energy \cite{ekli} and so on. At present, the categories of these theories are most widely accepted, playing a central role in the explanation of the anomalies in observed galactic rotation. There are three kinds of main theories as follows. (i) Dark matter theory. The dark matter is still one unsolved puzzle in essence. According to the moving velocity refer to the particles, dark matter can be approximately divided into three classes: cold, warm, and hot dark matter. In the dark matter theory, the most widely discussed model is based on the cold dark matter hypothesis, and the corresponding particle is one Weakly Interacting Massive Particle \cite{stadnik} . (ii) Grand unification theory. The axion \cite{klimchitskaya} is one hypothetical elementary particle, trying to figure out the strong CP problem in the Quantum Chromodynamics. A part of axions may possess a low mass within a specific range, forming an additional structure at the galactic scale, and attempting to explain the dark matter puzzle. (iii) Supersymmetric theory. The Supersymmetric theory is the foundation of Supergravity theory and of Superstring theory. Its hypothetical particle, sterile neutrino \cite{yhli} , is not able to interact via any of the fundamental interactions except gravity, and is one candidate of dark matter particles. Moreover a few scholars applied the quaternion and octonion to study various properties of dark matters.

J. C. Maxwell was the first to use the quaternion to study the physics properties of electromagnetic field. Undoubtedly it inspires other scholars to apply the quaternion and octonion to explore the electromagnetic and gravitational theories, and even the field theory relevant to the dark matter. In some described theories, the coordinates of quaternion or of octonion can even be complex numbers  \cite{weng1} . At the moment, they are called as the complex quaternion or octonion respectively.

A few scholars study the physics properties of electromagnetic and gravitational fields by means of the complex quaternion and octonion. H. T. Anastassiu $et~al$. explored the electromagnetic feature with the complex quaternion \cite{anastassiu} . K. Morita studied the quaternion field theory \cite{morita} . F. A. Doria used the complex quaternion to depict the gravitational theory \cite{doria} . V. Majernik deduced the modified Maxwell-like gravitational field equations with the complex quaternion \cite{majernik1} . M. Gogberashvili applied the complex octonion to research the electromagnetic field equations \cite{gogberashvili} . V. L. Mironov $et~al$. described the electromagnetic field equations and related features by the algebra of octonions \cite{mironov} . S. Demir made use of the octonion to discuss the gravitational field equations and relevant properties \cite{demir} .

Further a part of scholars explore the physics properties of dark matter by means of the quaternion and octonion. S. P. Brumby $et~al$. presented some global consequences of a model quaternionic quantum field theory, making the quaternionic structure a dynamical quantity naturally leads to the prediction of cosmic strings and non-baryonic hot dark matter candidates \cite{brumby1} . Taking the complex nature of quantum mechanics, which observed as a low energy effect of a broken quaternionic theory, the authors \cite{brumby2} explored the possibility that dark matter may arise as a consequence of the underlying quaternionic structure to the universe. V. Majernik \cite{majernik2} described a model of the universe consisting of a mixture of the ordinary matter and a so-called cosmic quaternionic field, investigating the interaction of the quaternionic field with the ordinary dark matter, and exerting a force on the moving dark matter which might possible create the dark matter in the early universe. S. Furui expressed equivalently a Dirac fermion as a 4-component spinor \cite{urui1} , which is a combination of two quaternions. And the latter can be treated as an octonion. The octonion possesses the triality symmetry, which relates three sets of spinors and two sets of vectors. If the electromagnetic interaction is sensitive to the triality symmetry (i.e. EM probe selects one triality sector), electromagnetic signals from the 5-transformed world would not be detected, and be treated as the dark matter. Reconstructing a left-handed fermion and a right-handed fermion in terms of the quaternion in the Yang-Mills Lagrangian, the author \cite{urui2} discussed the axial anomaly and the triality symmetry of octonion. According to the triality selection rules of octonions, the Pauli spinor is treated as a quaternion, and the Dirac spinor is treated as an octonion. Therefore the author \cite{urui3} assumed the dark matter may be able to be interpreted as matter emitting photons in a different triality sector, rather than that of electromagnetic probes in the world. B. C. Chanyal $et~al$. \cite{chanyal} analyzed the role of octonions in various unified field theories associated with the dyon and dark matter, reconstructing the field equations of hot and cold dark matter by means of split octonions.

\subsection{Limitation}

Making a detailed comparison and analysis of preceding studies, a few primal problems of these theories are found as follows. (1) Common origin. Up to now, the existing theories have not unified the dark matter and ordinary matter into one single field theory, and not found the common origin of dark matters and ordinary matters yet. However in the paper, the electromagnetic field accompanies with one adjoint-field (in Section 3), which feature is partly similar to that of gravitational field. If the electromagnetic adjoint-field is chosen as the dark matter field, the dark matter and ordinary matter possess the same integrating function of field potential. In other words, for the complex octonion integrating function of field potential, the dark matter and ordinary matter have one common origin, simplifying the theoretical model of dark matter significantly. (2) Additional contribution. So far the existing theories can not deduce the contribution of dark matter to the field equations, linear momentum, angular momentum, energy, torque, force, and power and so forth. But in the paper, the emergence of the dark matter is accompanied by the existence of electromagnetic field. In case there are both electromagnetic and gravitational fields, it is capable of inferring the field theory related to the dark matter, measuring and calculating the additional contribution of dark matter field in the complex quaternion space. (3) Particle category. The existing theories are not able to determine the particle category of dark matter until now. In the paper, the dark matter particle is independent of the ordinary matter particle. And it is able to predict that the dark matter has only three sorts of particles, for the double-source particles. The first sort of dark matter particle carries the mass; the second carries the electric charge; the third carries neither the mass nor the electric charge.

Making use of the complex octonion space and the complex quaternion operator, it is propitious to describe the physics properties of electromagnetic and gravitational fields. When the operator is extended from the complex quaternion operator to the complex octonion operator, the paper is appropriate for exploring not only the electromagnetic and gravitational fields, but also the electromagnetic and gravitational adjoint-fields. The electromagnetic adjoint-field possesses a part of physics properties of gravitational field, and can be chosen as one competitive candidate of dark matter field.

\section{Octonion space}

In the complex quaternion space $\mathbb{H}_g$ for the gravitational field, the basis vector is $\emph{\textbf{i}}_j$, the coordinates are $i r_0$ and $r_k$ . The quaternion radius vector is, $\mathbb{R}_g = i \emph{\textbf{i}}_0 r_0 + \Sigma \emph{\textbf{i}}_k r_k$. Similarly, in the complex $S$-quaternion (short for the second quaternion) space $\mathbb{H}_e$ for the electromagnetic field, the basis vector is $\emph{\textbf{I}}_j$ , the coordinates are $i R_0$ and $R_k$. The $S$-quaternion radius vector is, $\mathbb{R}_e = i \emph{\textbf{I}}_0 R_0 + \Sigma \emph{\textbf{I}}_k R_k$ . Herein $r_j$ and $R_j$ are all real. $r_0 = v_0 t$ , $t$ stands for the time, $v_0$ is the speed of light. $i$ is the imaginary unit. The symbol $\circ$ denotes the octonion multiplication. $\emph{\textbf{i}}_0 = 1$. $\emph{\textbf{i}}_k^2 = -1$. $\emph{\textbf{I}}_j^2 = -1$. $\emph{\textbf{I}}_j = \emph{\textbf{i}}_j \circ \emph{\textbf{I}}_0$ . $j = 0, 1, 2, 3$; $k = 1, 2, 3$.

Two independent complex quaternion spaces, $\mathbb{H}_g$ and $\mathbb{H}_e$ , are orthogonal to each other. They are able to combine together to become one complex octonion space $\mathbb{O}$ (Appendix A). In the complex octonion space $\mathbb{O}$ , the basis vectors are $\emph{\textbf{i}}_j$ and $\emph{\textbf{I}}_j$ , the coordinates are $i r_0$ , $r_k$ , $i k_{eg} R_0$ , and $k_{eg} R_k$ . The octonion radius vector is, $\mathbb{R} = \mathbb{R}_g + k_{eg} \mathbb{R}_e$ . Herein $k_{eg}$ is the coefficient.

In the complex quaternion space $\mathbb{H}_g$ , the quaternion operator is, $\square_g = i \partial_{r0} + \Sigma \emph{\textbf{i}}_k \partial_{rk}$ . In the complex $S$-quaternion space $\mathbb{H}_e$ , the $S$-quaternion operator is, $\square_e = i \emph{\textbf{I}}_0 \partial_{R0} + \Sigma \emph{\textbf{I}}_k \partial_{Rk}$ . Obviously, in the complex octonion space $\mathbb{O}$ , the octonion operator is, $\square = \square_g + k_{eg}^{-1} \square_e$ . Herein $\partial_{rj} = \partial / \partial r_j$ , $\partial_{Rj} = \partial / \partial R_j$ .

\section{Field and adjoint-field}

In the complex octonion space, making use of the complex quaternion operator and the physics quantity of gravitational field, it is able to yield the gravitational theory in the complex quaternion space. The gravitational theory can be applied to account for the phenomenon of astrophysics jets in the astronomy \cite{weng2} , including thrust, bipolarity, precession, collimation, continuing acceleration and so forth. Under certain approximate conditions, the gravitational theory will be able to be degenerated into the Newtonian gravitational theory. Following the same idea, by means of the complex quaternion operator and the physics quantity of electromagnetic field, it is capable of deducing the electromagnetic theory in the complex $S$-quaternion space. The electromagnetic theory is identical with the Maxwellian electromagnetic theory.

In the complex quaternion space $\mathbb{H}_g$ , the quaternion integrating function of field potential is, $\mathbb{X}_g = i x_0 + \Sigma \emph{\textbf{i}}_k x_k$ . Similarly, in the complex $S$-quaternion space $\mathbb{H}_e$, the $S$-quaternion integrating function of field potential is, $\mathbb{X}_e = i \emph{\textbf{I}}_0 X_0 + \Sigma \emph{\textbf{I}}_k X_k$. They can combine together to become one octonion integrating function of field potential, $\mathbb{X} = \mathbb{X}_g + k_{eg} \mathbb{X}_e $ , in the complex octonion space $\mathbb{O}$ . Herein $x_j$ and $X_j$ are all real.

When the quaternion operator, $\square_g$ , acts on the quaternion integrating function of field potential, $\mathbb{X}_g$ , it is able to obtain the quaternion field potential, $\mathbb{A}_g^g = i \square_g^\times \circ \mathbb{X}_g$ , of the gravitational field in the complex quaternion space $\mathbb{H}_g$ . Similarly, when the operator, $\square_g$ , acts on the $S$-quaternion integrating function of field potential, $\mathbb{X}_e$ , it is capable of achieving the $S$-quaternion field potential, $\mathbb{A}_e^g = i \square_g^\times \circ \mathbb{X}_e$ , of the electromagnetic field in the complex $S$-quaternion space $\mathbb{H}_e$ (Table 1).

Further in the complex octonion space $\mathbb{O}$ , the complex octonion operator, $\square$ , consists of the quaternion operator, $\square_g$ , and the $S$-quaternion operator, $\square_e$ . When the complex octonion operator, $\square$ , acts on the octonion integrating function of field potential $\mathbb{X}$ , it will generate four kinds of quaternion field potentials. The first two field potentials relate to the operator $\square_g$ , including the gravitational potential and electromagnetic potentials. The rest of them are relevant to the operator $\square_e$ , and are the potentials of two new `similar fields'. The new `similar field' is called as the adjoint-field temporarily.

For one fundamental field (gravitational or electromagnetic field), its adjoint-field is extended, accessorial, and accompanied. With the emergence of the fundamental field, the adjoint-field may burst out. In general, some physics properties of adjoint-field is similar to that of fundamental field to a certain extent.

When the $S$-quaternion operator, $\square_e$ , acts on the quaternion integrating function of field potential, $\mathbb{X}_g$ , it is capable of obtaining the field potential, $\mathbb{A}_g^e = i \square_e^\times \circ \mathbb{X}_g$, of the gravitational adjoint-field (G adjoint-field for short). It should be noted that the field potential, $\mathbb{A}_g^e$ , will be situated on the complex $S$-quaternion space $\mathbb{H}_e$ , according to the octonion multiplication. Similarly when the operator $\square_e$ , acts on the $S$-quaternion integrating function of field potential, $\mathbb{X}_e$ , it is able to achieve the field potential, $\mathbb{A}_e^e = i \square_e^\times \circ \mathbb{X}_e$ , of electromagnetic adjoint-field (E adjoint-field for short). And it stays at the complex quaternion space $\mathbb{H}_g$ .

In the complex octonion space $\mathbb{O}$ , the octonion field potential is defined as,
\begin{eqnarray}
\mathbb{A} = i \square^\times \circ \mathbb{X} ~,
\end{eqnarray}
where $\mathbb{A} = \mathbb{A}_g + k_{eg} \mathbb{A}_e$ . $\mathbb{A}_g = \mathbb{A}_g^g + \mathbb{A}_e^e$ , $\mathbb{A}_e = \mathbb{A}_e^g + k_{eg}^{-2} \mathbb{A}_g^e$ . $\mathbb{A}_g$ is the component of the octonion field potential, $\mathbb{A}$ , in the complex quaternion space $\mathbb{H}_g$ . And $\mathbb{A}_e$ is the component of the octonion field potential, $\mathbb{A}$ , in the complex $S$-quaternion space $\mathbb{H}_e$ . $\mathbb{A}_g^g = i A_{g0}^g + \Sigma \emph{\textbf{i}}_k A_{gk}^g$ , $\mathbb{A}_g^e = i \emph{\textbf{I}}_0 A_{g0}^e + \Sigma \emph{\textbf{I}}_k A_{gk}^e$ . $\mathbb{A}_e^g = i \emph{\textbf{I}}_0 A_{e0}^g + \Sigma \emph{\textbf{I}}_k A_{ek}^g$, $\mathbb{A}_e^e = i A_{e0}^e + \Sigma \emph{\textbf{i}}_k A_{ek}^e$ . $A_{gj}^g$ , $A_{gj}^e$ , $A_{ej}^g$ , and $A_{ej}^e$ are all real.

When the contribution of the adjoint-field can be neglected, the terms, $( A_{g0}^g / v_0 )$ and $( \Sigma \emph{\textbf{i}}_k A_{gk}^g )$, are respectively the scalar potential and vector potential of gravitational field in the gravitational theory. And the terms, $( \emph{\textbf{I}}_0 A_{e0}^g / v_0 )$ and $( \Sigma \emph{\textbf{I}}_k A_{ek}^g )$, are respectively the `scalar potential' and `vector potential' of electromagnetic field in the classical field theory. However, when there are adjoint-fields, two field potentials, $\mathbb{A}_g^g$ and $\mathbb{A}_e^e$ , will be coexistent, and their properties are similar partly. These two field potentials are daunting to distinguish to a certain extent. Also two field potentials, $\mathbb{A}_e^g$ and $\mathbb{A}_g^e$ , have similar properties partly, and are difficult to distinguish to a certain extent. In this case, it is necessary to consider the contribution of the adjoint-field, $\mathbb{A}_e^e$ and $\mathbb{A}_g^e$ , the gravitational potential will be extended into the term $\mathbb{A}_g$ , and the electromagnetic potential will be extended into the term $\mathbb{A}_e$ .

For the sake of convenience in the subsequent context, one can adopt the gauge condition for the integrating function of field potential, in Eq.(1). That is, the curl of vector part of the integrating function of field potential is chosen to be zero.
\begin{eqnarray}
\nabla_r \times \textbf{x} + \nabla_R \times \textbf{X} = 0 ~,~~  \nabla_r \times \textbf{X} + k_{eg}^{-2} \nabla_R \times \textbf{x} = 0 ~,
\end{eqnarray}
where $\nabla_r = \Sigma \emph{\textbf{i}}_k \partial_{rk}$ , $\nabla_R = \Sigma \emph{\textbf{I}}_k \partial_{Rk}$ . $\textbf{x} = \Sigma \emph{\textbf{i}}_k x_k$ , $\textbf{X} = \Sigma \emph{\textbf{I}}_k X_k$ .

Further the gravitational potential $\mathbb{A}_g$ and electromagnetic potential $\mathbb{A}_e$ are,
\begin{eqnarray}
\mathbb{A}_g = i A_{g0} + \Sigma \emph{\textbf{i}}_k A_{gk} ~,~~  \mathbb{A}_e = i \emph{\textbf{I}}_0 A_{e0} + \Sigma \emph{\textbf{I}}_k A_{ek} ~,
\end{eqnarray}
where $A_{gj}$ and $A_{ej}$ are all real.

The above states that it is able to obtain the G adjoint-field in the complex $S$-quaternion space, when the complex $S$-quaternion operator acts on the physics quantities of gravitational field in the complex quaternion space. The G adjoint-field can be combined with the electromagnetic field, exerting an influence on the ordinary field source (electric charge and current) and G adjoint-field source, and therefore impacting the movements of electric charge and current. When the complex $S$-quaternion operator acts on the physics quantities of electromagnetic field in the complex $S$-quaternion space, it is capable of yielding the E adjoint-field in the complex quaternion space. The E adjoint-field can be combined with the gravitational field, to exert an influence on the ordinary field source (mass and linear momentum) and E adjoint-field source, and then impacting the movements of mass and linear momentum. Subsequently the E adjoint-field can be chosen as one candidate of dark matter field.

In the field theory described with the complex octonion space and operator, two fundamental fields and two adjoint-fields are similar to the four surfaces of one tetrahedral. They are independent of each other, and constitute one whole together. We must investigate these `four fields' of the whole simultaneously, if we want to explore their physics properties profoundly.

\begin{table}[ph]
\tbl{The multiplication of the operator and octonion quantity.}
{\begin{tabular}{@{}ll@{}}
\hline
definition                    &  expression                                                                                                                             \\
\hline
$\nabla_r \cdot \textbf{a}$   &  $-(\partial_{r1} a_1 + \partial_{r2} a_2 + \partial_{r3} a_3)$                                                                         \\
$\nabla_r \times \textbf{a}$  &  $\emph{\textbf{i}}_1 ( \partial_{r2} a_3  - \partial_{r3} a_2 ) + \emph{\textbf{i}}_2 ( \partial_{r3} a_1 - \partial_{r1} a_3 )
                                 + \emph{\textbf{i}}_3 ( \partial_{r1} a_2 - \partial_{r2} a_1 )$                                                                       \\
$\nabla_r a_0$                &  $\emph{\textbf{i}}_1 \partial_{r1} a_0 + \emph{\textbf{i}}_2 \partial_{r2} a_0 + \emph{\textbf{i}}_3 \partial_{r3} a_0  $              \\
$\partial_{r0} \textbf{a}$    &  $\emph{\textbf{i}}_1 \partial_{r0} a_1 + \emph{\textbf{i}}_2 \partial_{r0} a_2 + \emph{\textbf{i}}_3 \partial_{r0} a_3  $              \\
\hline
$\nabla_r \cdot \textbf{A}$   &  $- \emph{\textbf{I}}_0 (\partial_{r1} A_1 + \partial_{r2} A_2 + \partial_{r3} A_3)  $                                                  \\
$\nabla_r \times \textbf{A}$  &  $- \emph{\textbf{I}}_1 ( \partial_{r2} A_3 - \partial_{r3} A_2 ) - \emph{\textbf{I}}_2 ( \partial_{r3} A_1 - \partial_{r1} A_3 )
                                 - \emph{\textbf{I}}_3 ( \partial_{r1} A_2 - \partial_{r2} A_1 )$                                                                       \\
$\nabla_r \circ \textbf{A}_0$ &  $\emph{\textbf{I}}_1 \partial_{r1} A_0 + \emph{\textbf{I}}_2 \partial_{r2} A_0  + \emph{\textbf{I}}_3 \partial_{r3} A_0  $             \\
$\partial_{r0} \textbf{A}$    &  $\emph{\textbf{I}}_1 \partial_{r0} A_1 + \emph{\textbf{I}}_2 \partial_{r0} A_2 + \emph{\textbf{I}}_3 \partial_{r0} A_3  $              \\
\hline
$\nabla_R \cdot \textbf{a}$   &  $\emph{\textbf{I}}_0 (\partial_{R1} a_1 + \partial_{R2} a_2 + \partial_{R3} a_3)$                                                      \\
$\nabla_R \times \textbf{a}$  &  $- \emph{\textbf{I}}_1 ( \partial_{R2} a_3 - \partial_{R3} a_2 ) - \emph{\textbf{I}}_2 ( \partial_{R3} a_1 - \partial_{R1} a_3 )
                                 - \emph{\textbf{I}}_3 ( \partial_{R1} a_2 - \partial_{R2} a_1 )$                                                                       \\
$\nabla_R a_0$                &  $\emph{\textbf{I}}_1 \partial_{R1} a_0 + \emph{\textbf{I}}_2 \partial_{R2} a_0 + \emph{\textbf{I}}_3 \partial_{R3} a_0  $              \\
$\emph{\textbf{I}}_0 \partial_{R0} \circ \textbf{a}$     &  $- \emph{\textbf{I}}_1 \partial_{R0} a_1 - \emph{\textbf{I}}_2 \partial_{R0} a_2
                                 - \emph{\textbf{I}}_3 \partial_{R0} a_3  $                                                                                             \\
\hline
$\nabla_R \cdot \textbf{A}$   &  $-(\partial_{R1} A_1 + \partial_{R2} A_2 + \partial_{R3} A_3) $                                                                        \\
$\nabla_R \times \textbf{A}$  &  $-\emph{\textbf{i}}_1 ( \partial_{R2} A_3 - \partial_{R3} A_2 ) - \emph{\textbf{i}}_2 ( \partial_{R3} A_1 - \partial_{R1} A_3 )
                                 - \emph{\textbf{i}}_3 ( \partial_{R1} A_2 - \partial_{R2} A_1 )$                                                                       \\
$\nabla_R \circ \textbf{A}_0$ &  $- \emph{\textbf{i}}_1 \partial_{R1} A_0 - \emph{\textbf{i}}_2 \partial_{R2} A_0 - \emph{\textbf{i}}_3 \partial_{R3} A_0  $            \\
$\emph{\textbf{I}}_0 \partial_{R0} \circ \textbf{A}$     &  $\emph{\textbf{i}}_1 \partial_{R0} A_1 + \emph{\textbf{i}}_2 \partial_{R0} A_2
                                 + \emph{\textbf{i}}_3 \partial_{R0} A_3  $                                                                                             \\
\hline
\end{tabular}}
\end{table}

\section{Field strength}

In the complex octonion space $\mathbb{O}$ , when the adjoint-field is not equal to zero, it is able to obtain the gravitational strength $\mathbb{F}_g^g$ in the complex quaternion space, by means of the quaternion operator and the component $\mathbb{A}_g$ of octonion field potential. Also it can achieve the electromagnetic strength $\mathbb{F}_e^g$ in the complex $S$-quaternion space, making use of the quaternion operator and the component $\mathbb{A}_e$ of octonion field potential. Further it may acquire the G adjoint-field strength $\mathbb{F}_g^e$ in the complex $S$-quaternion space, taking advantage of the $S$-quaternion operator and the component $\mathbb{A}_g$ of octonion field potential. And it is capable of attaining the E adjoint-field strength $\mathbb{F}_e^e$ in the complex quaternion space, from the $S$-quaternion operator and the component $\mathbb{A}_e$ of octonion field potential.

In the complex quaternion space $\mathbb{H}_g$ , when the quaternion operator $\square_g$ acts on the component $\mathbb{A}_g$ of octonion field potential, it is able to deduce the quaternion gravitational strength, $\mathbb{F}_g^g = \square_g \circ \mathbb{A}_g$ . Expanding of this equation yields the components of gravitational strength. Under certain approximate conditions, they will be degenerated into the gravitational acceleration and the gravitational component $\textbf{b}$ in the gravitational theory. Similarly in the complex $S$-quaternion space $\mathbb{H}_e$ , when the operator $\square_g$ acts on the component $\mathbb{A}_e$ of octonion field potential, it is capable of inferring the $S$-quaternion electromagnetic strength, $\mathbb{F}_e^g = \square_g \circ \mathbb{A}_e$ . Expanding of this equation produces the components of electromagnetic strength. Under certain approximate conditions, they will be degenerated into the electric field intensity and the magnetic flux density in the classical electromagnetic theory (Table 2).

When the $S$-quaternion operator $\square_e$ acts on the component $\mathbb{A}_g$ of octonion field potential, it is able to deduce the G adjoint-field strength, $\mathbb{F}_g^e = \square_e \circ \mathbb{A}_g$ . It should be noted that the G adjoint-field strength $\mathbb{F}_g^e$ will be situated on the $S$-quaternion space $\mathbb{H}_e$ , according to the octonion multiplication. Similarly, when the operator $\square_e$ acts on the component $\mathbb{A}_e$ of octonion field potential, it is capable of obtaining the E adjoint-field strength, $\mathbb{F}_e^e = \square_e \circ \mathbb{A}_e$ , which will remain in the quaternion space $\mathbb{H}_g$ . (Appendix B)

In the complex octonion space $\mathbb{O}$ , the octonion field strength is defined as,
\begin{eqnarray}
\mathbb{F} = \square \circ \mathbb{A}  ~,
\end{eqnarray}
where $\mathbb{F} = \mathbb{F}_g + k_{eg} \mathbb{F}_e$ . $\mathbb{F}_g = \mathbb{F}_g^g + \mathbb{F}_e^e$ , $\mathbb{F}_e = \mathbb{F}_e^g + k_{eg}^{-2} \mathbb{F}_g^e$ . $\mathbb{F}_g$ is the component of octonion field strength $\mathbb{F}$ in the quaternion space $\mathbb{H}_g$ . And $\mathbb{F}_e$ is the component of octonion field strength $\mathbb{F}$ in the $S$-quaternion space $\mathbb{H}_e$ . $\mathbb{F}_g^g = i F_{g0}^g + \Sigma \emph{\textbf{i}}_k F_{gk}^g$ , $\mathbb{F}_g^e = i \emph{\textbf{I}}_0 F_{g0}^e + \Sigma \emph{\textbf{I}}_k F_{gk}^e$ . $\mathbb{F}_e^g = i \emph{\textbf{I}}_0 F_{e0}^g + \Sigma \emph{\textbf{I}}_k F_{ek}^g$ , $\mathbb{F}_e^e = i F_{e0}^e + \Sigma \emph{\textbf{i}}_k F_{ek}^e$ . $F_{g0}^g$ , $F_{g0}^e$ , $F_{e0}^g$ , and $F_{e0}^e$ are all real. $F_{gk}^g$ , $F_{gk}^e$ , $F_{ek}^g$ , and $F_{ek}^e$ are all complex numbers.

When there are adjoint-fields, two field strengths, $\mathbb{F}_g^g$ and $\mathbb{F}_e^e$ , will be coexistent, and their properties are similar partly. These two field strengths are tough to distinguish to a certain extent. Also two field strengths, $\mathbb{F}_e^g$ and $\mathbb{F}_g^e$ , have similar properties partly, and are difficult to distinguish to a certain extent. Therefore in the case that it is necessary to consider the contribution of field strengths, $\mathbb{F}_e^e$ and $\mathbb{F}_g^e$ , the gravitational strength will be extended into the term $\mathbb{F}_g$ , and the electromagnetic strength will be extended into the term $\mathbb{F}_e$ .

For the sake of convenience in the subsequent context, one may apply the gauge condition for the field potential, in Eq.(4). That is, the scalar part of the field strength is chosen to be equal to zero.
\begin{eqnarray}
F_{g0} = F_{g0}^g + F_{e0}^e = 0 ~,~~  F_{e0} = F_{g0}^e + k_{eg}^{-2} F_{e0}^g = 0 ~,
\end{eqnarray}
where $\mathbb{F}_g = F_{g0} + \Sigma \emph{\textbf{i}}_k F_{gk}$ , $\mathbb{F}_e = \emph{\textbf{I}}_0 F_{e0} + \Sigma \emph{\textbf{I}}_k F_{ek}$ . $F_{g0}$ and $F_{e0}$ are all real. $F_{gk}$ and $F_{ek}$ are all complex numbers.

As a result, the gravitational strength $\mathbb{F}_g$ and electromagnetic strength $\mathbb{F}_e$ are,
\begin{eqnarray}
\mathbb{F}_g = i \textbf{g} / v_0 + \textbf{b} ~,~~  \mathbb{F}_e = i \textbf{E} / v_0 + \textbf{B} ~,
\end{eqnarray}
where the gravitational acceleration is, $\textbf{g} = \Sigma \emph{\textbf{i}}_k g_k$ , and the component $\textbf{b}$ of gravitational strength is, $\textbf{b} = \Sigma \emph{\textbf{i}}_k b_k$ . The electric field intensity is, $\textbf{E} = \Sigma \emph{\textbf{I}}_k E_k$ , and the magnetic flux density is, $\textbf{B} = \Sigma \emph{\textbf{I}}_k B_k$ . $g_k$ , $b_k$ , $E_k$ , and $B_k$ are all real.

The above means that it is able to deduce not only the gravitational strength and electromagnetic strength, but also the G adjoint-field strength and E adjoint-field strength, making use of the octonion operator and field potential. Obviously the gravitational potential and E adjoint-field potential are mixed together. And they are difficult to distinguish sometimes. From the definition, $\mathbb{F}_g^g = \square_g \circ ( \mathbb{A}_g^g + \mathbb{A}_e^e )$, it is easy to find that it is still able to produce the gravitational strength $\mathbb{F}_g^g$ , from the E adjoint-field potential $\mathbb{A}_e^e$ , even if the gravitational potential $\mathbb{A}_g^g$ is equal to zero. Moreover, from the definition, $\mathbb{F}_g^e = \square_e \circ ( \mathbb{A}_g^g + \mathbb{A}_e^e )$, it is still capable of deducing the G adjoint-field strength $\mathbb{F}_g^e$ , from the E adjoint-field potential $\mathbb{A}_e^e$ , even if the gravitational potential $\mathbb{A}_g^g$ is equal to zero (Table 3).

Similarly the electromagnetic potential and G adjoint-field potential are mixed together. From the definition, $\mathbb{F}_e^g = \square_g \circ ( \mathbb{A}_e^g + k_{eg}^{-2} \mathbb{A}_g^e )$, it is easy to find that it is still capable of deducing the electromagnetic strength $\mathbb{F}_e^g$ , from the G adjoint-field potential $\mathbb{A}_g^e$ , even if the electromagnetic potential $\mathbb{A}_e^g$ is equal to zero. Moreover, from the definition, $\mathbb{F}_e^e = \square_e \circ ( \mathbb{A}_e^g + k_{eg}^{-2} \mathbb{A}_g^e )$, it is still capable of inferring the E adjoint-field strength $\mathbb{F}_e^e$ , from the G adjoint-field potential $\mathbb{A}_g^e$ , even if the electromagnetic potential $\mathbb{A}_e^g$ is equal to zero.

In the subsequent context, the physics quantity of gravitational field will mix with that of E adjoint-field closely, including the field potential, field strength, and field source and so forth. By all appearances, it is tough to distinguish two field strengths, $\mathbb{F}_g^g$ and $\mathbb{F}_e^e$ , in general. Only in an extreme condition, it may be possible to appraise the interference of term $\mathbb{F}_e^e$ on $\mathbb{F}_g^g$ . In a similar way, the physics quantity of electromagnetic field will mix with that of G adjoint-field closely, including the field potential, field strength, and field source and so forth. In general, it is hard to discriminate two field strengths, $\mathbb{F}_e^g$ and $\mathbb{F}_g^e$ . Only in an extreme condition, it may be possible to assess the interference of term $\mathbb{F}_g^e$ on $\mathbb{F}_e^g$ .

\section{Field source}

In the complex octonion space $\mathbb{O}$ , from the quaternion operator and the component $\mathbb{F}_g$ of octonion field strength, it is able to obtain the gravitational source $\mathbb{S}_g^g$ in the complex quaternion space. And it is capable of achieving the electromagnetic source $\mathbb{S}_e^g$ in the complex $S$-quaternion space, by means of the quaternion operator and the component $\mathbb{F}_e$ of octonion field strength. Further, it is able to infer the G adjoint-filed source $\mathbb{S}_g^e$ in the complex $S$-quaternion space, making use of the $S$-quaternion operator and the component $\mathbb{F}_g$ of octonion field strength. In terms of the $S$-quaternion operator and the component $\mathbb{F}_e$ , it is capable of deducing the E adjoint-filed source $\mathbb{S}_e^e$ in the complex quaternion space (Table 4).

When the quaternion operator $\square_g$ acts on the component $\mathbb{F}_g$ of octonion field strength, it is able to deduce the quaternion gravitational source, $\mu_g^g \mathbb{S}_g^g = - \square_g^* \circ \mathbb{F}_g$ , in the complex quaternion space $\mathbb{H}_g$ . Expanding of this equation yields the gravitational field equations, which can be degenerated into the Newton's law of universal gravitation, under certain approximate conditions (Appendix C). When the operator $\square_g$ acts on the component $\mathbb{F}_e$ of octonion field strength, it is capable of inferring the $S$-quaternion electromagnetic source, $\mu_e^g \mathbb{S}_e^g = - \square_g^* \circ \mathbb{F}_e $. Expanding of this equation produces the electromagnetic field equations, which can be degenerated into the Maxwell's electromagnetic field equations, under certain approximate conditions. Herein $\mu_g^g$ and $\mu_e^g$ both are coefficients. $\mu_g^g < 0$, and $\mu_e^g > 0$.

Further, when the $S$-quaternion operator $\square_e$ acts on the component $\mathbb{F}_g$ of octonion field strength, it is able to deduce the G adjoint-field source, $\mu_g^e \mathbb{S}_g^e = - \square_e^* \circ \mathbb{F}_g$. It should be noted that the G adjoint-field source $\mathbb{S}_g^e$ will be situated on the $S$-quaternion space $\mathbb{H}_e$ , according to the octonion multiplication. Similarly, when the operator $\square_e$ acts on the component $\mathbb{F}_e$ of octonion field strength, it is capable of obtaining the E adjoint-field source, $\mu_e^e \mathbb{S}_e^e = - \square_e^* \circ \mathbb{F}_e$ . And the E adjoint-field source $\mathbb{S}_e^e$ remains in the quaternion space $\mathbb{H}_g$ . Herein $\mu_g^e$ and $\mu_e^e$ both are coefficients.

In the complex octonion space $\mathbb{O}$ , the octonion field source is written as,
\begin{eqnarray}
\mu \mathbb{S} = - ( i \mathbb{F} / v_0 + \square )^* \circ \mathbb{F}
= \mu_g \mathbb{S}_g + k_{eg} \mu_e \mathbb{S}_e - i \mathbb{F}^* \circ \mathbb{F} / v_0 ~,
\end{eqnarray}
where $\ast$ indicates the conjugation of octonion. $\mu_g \mathbb{S}_g = \mu_g^g \mathbb{S}_g^g + \mu_e^e \mathbb{S}_e^e$ , $\mu_e \mathbb{S}_e = \mu_e^g \mathbb{S}_e^g + k_{eg}^{-2} \mu_g^e \mathbb{S}_g^e$ . The term $\mathbb{S}_g$ is one part of component of octonion field source $\mathbb{S}$ in the complex quaternion space $\mathbb{H}_g$ . And $\mathbb{S}_e$ is the component of octonion field source $\mathbb{S}$ in the complex $S$-quaternion space $\mathbb{H}_e$ . $\mathbb{S}_g^g = i S_{g0}^g + \Sigma \emph{\textbf{i}}_k S_{gk}^g$ , $\mathbb{S}_g^e = i \emph{\textbf{I}}_0 S_{g0}^e + \Sigma \emph{\textbf{I}}_k S_{gk}^e$. $\mathbb{S}_e^g = i \emph{\textbf{I}}_0 S_{e0}^g + \Sigma \emph{\textbf{I}}_k S_{ek}^g$ , $\mathbb{S}_e^e = i S_{e0}^e + \Sigma \emph{\textbf{i}}_k S_{ek}^e$ . $S_{gj}^g$ , $S_{gj}^e$ , $S_{ej}^g$ , and $S_{ej}^e$ are all real. $\mu$, $\mu_g$ , and $\mu_e$ are coefficients.

When there are adjoint-fields, two field source, $\mathbb{S}_g^g$ and $\mathbb{S}_e^e$ , will be coexistent, and their features are partly similar. These two field sources are hard to distinguish to a certain extent. Also two field sources, $\mathbb{S}_e^g$ and $\mathbb{S}_g^e$ , have similar properties partly, and are difficult to be detached from each other to a certain extent. In this case, it is necessary to consider the contribution of field sources, $\mathbb{S}_e^e$ and $\mathbb{S}_g^e$ , the gravitational source will be extended into the term $\mathbb{S}_g$ , and the electromagnetic source will be extended into the term $\mathbb{S}_e$ .

Obviously the imaginary part, $\square_V$ , of the octonion operator $\square$ may form one operator also, which is called the velocity operator. Acting the velocity operator on the octonion radius vector, will yield the octonion velocity. That is,
\begin{eqnarray}
\mathbb{V} = v_0 \square_V \circ \mathbb{R} ~,
\end{eqnarray}
where $\square_V = \partial_{r0} + k_{eg}^{-1} \emph{\textbf{I}}_0 \partial_{R0}$ . $\mathbb{V} = \mathbb{V}_g^g + k_{eg} \mathbb{V}_e^g + k_{eg}^{-1} (\mathbb{V}_g^e + k_{eg} \mathbb{V}_e^e )$ . $\mathbb{V}_g^g = \partial_{r0} \mathbb{R}_g$, $\mathbb{V}_g^e = \emph{\textbf{I}}_0 \partial_{R0} \circ \mathbb{R}_g$ . $\mathbb{V}_e^g = \partial_{r0} \mathbb{R}_e$ , $\mathbb{V}_e^e = \emph{\textbf{I}}_0 \partial_{R0} \circ \mathbb{R}_e$ . $\mathbb{V}_g^g = i V_{g0}^g + \Sigma \emph{\textbf{i}}_k V_{gk}^g$ , $\mathbb{V}_g^e = i \emph{\textbf{I}}_0 V_{g0}^e + \Sigma \emph{\textbf{I}}_k V_{gk}^e$ . $\mathbb{V}_e^g = i \emph{\textbf{I}}_0 V_{e0}^g + \Sigma \emph{\textbf{I}}_k V_{ek}^g$ , $\mathbb{V}_e^e = i V_{e0}^e + \Sigma \emph{\textbf{i}}_k V_{ek}^e$ . $V_{g0}^g = v_0$ . $V_{gj}^g$ , $V_{gj}^e$, $V_{ej}^g$ , and $V_{ej}^e$ are all real.

For the single particle, there are the relationship among the field sources and the velocities.
\begin{eqnarray}
\mathbb{S}_g^g = m_g^g \mathbb{V}_g^g ~,~~  \mathbb{S}_g^e = m_g^e \mathbb{V}_g^e ~,~~  \mathbb{S}_e^g = m_e^g \mathbb{V}_e^g ~,~~  \mathbb{S}_e^e = m_e^e \mathbb{V}_e^e ~,
\end{eqnarray}
where $m_g^g$ is the mass density, $m_e^g$ is the density of electric charge , $m_g^e$ is the `charge' of G adjoint-field, and $m_e^e$ is the `charge' of G adjoint-field.

The above means that it is able to deduce not only the gravitational source and electromagnetic source, but also the G adjoint-field source and E adjoint-field source, making use of the octonion operator and field strength. Obviously the gravitational strength and E adjoint-field strength are mixed together. And they are difficult to distinguish sometimes. By means of the definition, $\mu_g^g \mathbb{S}_g^g = - \square_g^* \circ ( \mathbb{F}_g^g + \mathbb{F}_e^e )$, it is easy to find that it is able to produce the gravitational source $\mathbb{S}_g^g$ , from either the E adjoint-field strength $\mathbb{F}_e^e$ , or the gravitational strength $\mathbb{F}_g^g$ . In other words, the gravitational strength $\mathbb{F}_g^g$ and E adjoint-field strength $\mathbb{F}_e^e$ both make a contribution towards the gravitational source $\mathbb{S}_g^g$ . The measurement of the gravitational strength, $\mathbb{F}_g^g$ , is only able to determine one part of gravitational source $\mathbb{S}_g^g$ . Meanwhile the measurement of the E adjoint-field strength, $\mathbb{F}_e^e$ , is capable of deducing another part of gravitational source $\mathbb{S}_g^g$ . Moreover, in terms of the definition, $\mu_g^e \mathbb{S}_g^e = - \square_e^* \circ ( \mathbb{F}_g^g + \mathbb{F}_e^e )$, it is still capable of inferring the G adjoint-field source $\mathbb{S}_g^e$ , from the E adjoint-field strength $\mathbb{F}_e^e$ , even if the gravitational strength $\mathbb{F}_g^g$ is equal to zero. By means of measuring $\mathbb{F}_e^e$ and $\mathbb{F}_g^g$ simultaneously, it is possible to make sure the G adjoint-field source $\mathbb{S}_g^e$ (Table 5).

Similarly the electromagnetic strength and G adjoint-field strength are mixed together. From the definition, $\mu_e^g \mathbb{S}_e^g = - \square_g^* \circ ( \mathbb{F}_e^g + k_{eg}^{-2} \mathbb{F}_g^e )$, it is easy to find that it is capable of deducing the electromagnetic source $\mathbb{S}_e^g$ , from either the G adjoint-field strength $\mathbb{F}_g^e$ , or the electromagnetic strength $\mathbb{F}_e^g$ . In other words, the electromagnetic strength $\mathbb{F}_e^g$ and G adjoint-field strength $\mathbb{F}_g^e$ both make a contribution to the electromagnetic source $\mathbb{S}_e^g$ . The measurement of the electromagnetic strength $\mathbb{F}_e^g$ , is only able to infer one part of electromagnetic source $\mathbb{S}_e^g$ . Meanwhile the measurement of the G adjoint-field strength $\mathbb{F}_g^e$ , is capable of determining another part of electromagnetic source $\mathbb{S}_e^g$ . Moreover, from the definition, $\mu_e^e \mathbb{S}_e^e = - \square_e^* \circ ( \mathbb{F}_e^g + k_{eg}^{-2} \mathbb{F}_g^e )$, it is still capable of deducing the E adjoint-field source $\mathbb{S}_e^e$ , from the G adjoint-field strength $\mathbb{F}_g^e$ , even if the electromagnetic strength $\mathbb{F}_e^g$ is equal to zero. In terms of measuring $\mathbb{F}_g^e$ and $\mathbb{F}_g^e$ simultaneously, it may be possible to make certain the E adjoint-field source $\mathbb{S}_e^e$ .

\section{Angular momentum}

In the electromagnetic and gravitational theories described with the complex octonion, the definition of octonion angular momentum encompasses the angular momentum in the complex quaternion space $\mathbb{H}_g$ , and the electric dipolar moment and magnetic dipolar moment in the complex $S$-quaternion space $\mathbb{H}_e$ .

In the complex octonion space $\mathbb{O}$ , from the octonion field source $\mathbb{S}$ , it is able to define the octonion linear momentum as follows,
\begin{eqnarray}
\mathbb{P} = \mu \mathbb{S} / \mu_g^g ~,
\end{eqnarray}
where $\mathbb{P} = \mathbb{P}_g + k_{eg} \mathbb{P}_e$ . $\mathbb{P}_g = \mathbb{P}_g^g + \mathbb{P}_e^e$ , $\mathbb{P}_e = \mathbb{P}_e^g + k_{eg}^{-2} \mathbb{P}_g^e$ . $\mathbb{P}_g = i P_{g0} + \Sigma \emph{\textbf{i}}_k P_{gk}$ . $\mathbb{P}_e = i \emph{\textbf{I}}_0 P_{e0} + \Sigma \emph{\textbf{I}}_k P_{ek}$ . $\mathbb{P}_g$ is the component of the octonion linear momentum $\mathbb{P}$ in the complex quaternion space $\mathbb{H}_g$ , and $\mathbb{P}_e$ is the component of the octonion linear momentum $\mathbb{P}$ in the complex $S$-quaternion space $\mathbb{H}_e$ . $\mathbb{P}_g^g = \mathbb{S}_g^g - i \mathbb{F}^* \circ \mathbb{F} / ( \mu_g^g v_0 )$, $\mathbb{P}_g^e = \mu_g^e \mathbb{S}_g^e / \mu_g^g$ . $\mathbb{P}_e^g = \mu_e^g \mathbb{S}_e^g / \mu_g^g$ , $\mathbb{P}_e^e = \mu_e^e \mathbb{S}_e^e / \mu_g^g$ . $\mathbb{P}_g^g = i P_{g0}^g + \Sigma \emph{\textbf{i}}_k P_{gk}^g$ , $\mathbb{P}_g^e = i \emph{\textbf{I}}_0 P_{g0}^e + \Sigma \emph{\textbf{I}}_k P_{gk}^e$ . $\mathbb{P}_e^g = i \emph{\textbf{I}}_0 P_{e0}^g + \Sigma \emph{\textbf{I}}_k P_{ek}^g$ , $\mathbb{P}_e^e = i P_{e0}^e + \Sigma \emph{\textbf{i}}_k P_{ek}^e$ . $P_{gj}^g$ , $P_{gj}^e$ , $P_{ej}^g$ , and $P_{ej}^e$ are all real.

When there are adjoint-fields, two linear momenta, $\mathbb{P}_g^g$ and $\mathbb{P}_e^e$ , will be coexistent, and their properties are similar partly. Also the properties of two linear momenta, $\mathbb{P}_e^g$ and $\mathbb{P}_g^e$ , will be similar partly. The term $\mathbb{P}_e^e$ should be considered as one supplemental part, to enhance the influence of the $\mathbb{P}_g^g$ , and therefore form the component $\mathbb{P}_g$ of $\mathbb{P}$ . And the term $\mathbb{P}_g^e$ should be regarded as one supplemental part, to reinforce the influence of the $\mathbb{P}_e^g$ , and then form the component $\mathbb{P}_e$ of $\mathbb{P}$ . Consequently it is necessary to consider the contribution of linear momenta, $\mathbb{P}_e^e$ and $\mathbb{P}_g^e$ , the term $\mathbb{P}_g$ will be the quaternion linear momentum, and the term $( \mu_g^g \mathbb{P}_e / \mu_e )$ will be the $S$-quaternion electric current, in the electromagnetic and gravitational fields. $( P_{g0} / v_0 )$ and $( \Sigma \emph{\textbf{i}}_k P_{gk} )$ are respectively the density of gravitational mass and of linear momentum in the electromagnetic and gravitational fields. $( \emph{\textbf{I}}_0 P_{e0} / v_0 ) ( \mu_g^g / \mu_e )$ and $( \Sigma \emph{\textbf{I}}_k P_{ek} ) ( \mu_g^g / \mu_e )$ are the density of electric charge and of electric current in the electromagnetic and gravitational fields respectively.

In the complex octonion space $\mathbb{O}$ , the octonion angular momentum $\mathbb{L}$ can be defined from the octonion linear momentum $\mathbb{P}$ and radius vector $\mathbb{R}$ , that is,
\begin{eqnarray}
\mathbb{L} = ( \mathbb{R} + k_{rx} \mathbb{X} )^\times \circ \mathbb{P} ,
\end{eqnarray}
where $\mathbb{L} = \mathbb{L}_g + k_{eg} \mathbb{L}_e$ . $\mathbb{L}_g$ is the component of octonion angular momentum $\mathbb{L}$ in the complex quaternion space $\mathbb{H}_g$ , and $\mathbb{L}_e$ is the component of $\mathbb{L}$ in the complex $S$-quaternion space $\mathbb{H}_e$ . $k_{rx}$ is the coefficient. The symbol $\times$ denotes the complex conjugate.

In the complex quaternion space $\mathbb{H}_g$ , the component $\mathbb{L}_g$ of octonion angular momentum is,
\begin{eqnarray}
\mathbb{L}_g = \mathbb{L}_g^g + k_{eg}^2 \mathbb{L}_e^e ~,
\end{eqnarray}
where $\mathbb{L}_g^g = ( \mathbb{R}_g + k_{rx} \mathbb{X}_g )^\times \circ \mathbb{P}_g$ , $\mathbb{L}_e^e = ( \mathbb{R}_e + k_{rx} \mathbb{X}_e )^\times \circ \mathbb{P}_e$ . $\mathbb{R}_g^+ = \mathbb{R}_g + k_{rx} \mathbb{X}_g$ . $\mathbb{R}_e^+ = \mathbb{R}_e + k_{rx} \mathbb{X}_e$ . $r_j^+ = r_j + k_{rx} x_j$ , $R_j^+ = R_j + k_{rx} X_j$ . $\textbf{r}^+ = \Sigma r_k^+ \emph{\textbf{i}}_k$ . $\textbf{R}_0^+ = R_0^+ \emph{\textbf{I}}_0$, $\textbf{R}^+ = \Sigma R_k^+ \emph{\textbf{I}}_k$ . $\mathbb{P}_g = i p_0 + \textbf{p}$ . $p_0 = P_{g0}$ , $\textbf{p} = \Sigma \emph{\textbf{i}}_k P_{gk}$ . $\textbf{P}_e = i \textbf{P}_0 + \textbf{P}$ . $\textbf{P}_0 = \emph{\textbf{I}}_0 P_{e0}$, $\textbf{P} = \Sigma \emph{\textbf{I}}_k P_{ek}$ .

The above can be rewritten as,
\begin{eqnarray}
\mathbb{L}_g = L_{g0} + i \textbf{L}_g^i + \textbf{L}_g ,
\end{eqnarray}
where $L_{g0} = r_0^+ p_0 + \textbf{r}^+ \cdot \textbf{p} + k_{eg}^2 ( \textbf{R}_0^+ \circ \textbf{P}_0 + \textbf{R}^+ \cdot \textbf{P} )$ , $\textbf{L}_g^i = p_0 \textbf{r}^+ - r_0^+ \textbf{p}  + k_{eg}^2 ( \textbf{R}^+ \circ \textbf{P}_0 - \textbf{R}_0^+ \circ \textbf{P} )$ , $\textbf{L}_g = \textbf{r}^+ \times \textbf{p} + k_{eg}^2 \textbf{R}^+ \times \textbf{P} $ . $\textbf{L}_g = \Sigma L_{gk} \emph{\textbf{i}}_k$ . $\textbf{L}_g^i = \Sigma L^i_{gk} \emph{\textbf{i}}_k$ . $\textbf{L}_g$ is the angular momentum. $( k_{eg}^2 \textbf{R}^+ \times \textbf{P} )$ can be considered as the contribution of the E adjoint-field to the angular momentum. $k_{eg}^2 ( \textbf{R}_0^+ \circ \textbf{P}_0 + \textbf{R}^+ \cdot \textbf{P} )$ and $k_{eg}^2 ( \textbf{R}^+ \circ \textbf{P}_0 - \textbf{R}_0^+ \circ \textbf{P} )$ will be regarded as the contribution from the E adjoint-field. $L_{gj}$ and $L^i_{gk}$ are all real.

In the complex $S$-quaternion space $\mathbb{H}_e$ , the component $\mathbb{L}_e$ of octonion angular momentum is,
\begin{eqnarray}
\mathbb{L}_e = \mathbb{L}_e^g + \mathbb{L}_g^e ~,
\end{eqnarray}
where $\mathbb{L}_g^e = ( \mathbb{R}_e + k_{rx} \mathbb{X}_e )^\times \circ \mathbb{P}_g$ , $\mathbb{L}_e^g = ( \mathbb{R}_g + k_{rx} \mathbb{X}_g )^\times \circ \mathbb{P}_e$ .

The above can be rewritten as,
\begin{eqnarray}
\mathbb{L}_e = \textbf{L}_{e0} + i \textbf{L}_e^i + \textbf{L}_e ~,
\end{eqnarray}
where $\textbf{L}_{e0} = r_0^+ \textbf{P}_0 + \textbf{r}^+ \cdot \textbf{P} + p_0 \textbf{R}_0^+ + \textbf{R}^+ \cdot \textbf{p} $ , $\textbf{L}_e^i = - r_0^+ \textbf{P} + \textbf{r}^+ \circ \textbf{P}_0 - \textbf{R}_0^+ \circ \textbf{p} + p_0 \textbf{R}^+ $, $\textbf{L}_e = \textbf{r}^+ \times \textbf{P} + \textbf{R}^+ \times \textbf{p} $ . $\textbf{L}_{e0} = L_{e0} \emph{\textbf{I}}_0$ . $\textbf{L}_e = \Sigma L_{ek} \emph{\textbf{I}}_k$ . $\textbf{L}_e^i = \Sigma L^i_{ek} \emph{\textbf{I}}_k$. $\textbf{L}_e^i$ is the electric dipole moment. $( - \textbf{R}_0^+ \circ \textbf{p} + p_0 \textbf{R}^+ )$ can be considered as the contribution of the G adjoint-field to the electric dipole moment. $\textbf{L}_e$ is the magnetic dipole moment. $( \textbf{R}^+ \times \textbf{p} )$ will be regarded as the contribution of G adjoint-field to the magnetic dipole moment. $\textbf{L}_{e0}$ is called as the scalar-like magnetic moment temporarily, which is relevant to the spin magnetic moment. $( \textbf{R}^+ \cdot \textbf{p} )$ can be considered as the contribution of the G adjoint-field to the magnetic dipole moment. $L_{ej}$ and $L^i_{ek}$ are all real.

The above reveals that the definition of octonion angular momentum is able to cover some physics quantities, which were regarded as independent to each other, including the electric dipole moment, magnetic dipole moment, and angular momentum and so forth. In the electromagnetic and gravitational theories with the complex octonion, some physics quantities, which were considered as independent parts before, are actually consistent in essence.

Obviously the adjoint-field is one important component of the electromagnetic and gravitational fields described with the complex octonion. In the octonion space, it is necessary to consider the contribution of adjoint-field within certain limits. Under a few special situations, the contribution of adjoint-field may be significant enough. Such as, some movements relate with the E adjoint-field source may make a contribution to the angular momentum of galaxy and so forth.

\section{Octonion torque}	

The definition of octonion torque is able to contain the torque, work, and energy in the complex quaternion space $\mathbb{H}_g$ , and the derivative of electric dipole moment and of magnetic dipole moment in the complex $S$-quaternion space $\mathbb{H}_e$ , according to the electromagnetic and gravitational theories described with the complex octonion.

In the complex octonion space $\mathbb{O}$ , from the octonion operator $\square$ , octonion field strength $\mathbb{F}$ , and octonion angular momentum $\mathbb{L}$ , the octonion torque is defined as,
\begin{eqnarray}
\mathbb{W} = - v_0 ( i \mathbb{F} / v_0 + \square ) \circ \mathbb{L} ~,
\end{eqnarray}
where $\mathbb{W} = \mathbb{W}_g + k_{eg} \mathbb{W}_e$ . The term $\mathbb{W}_g$ is the component of octonion torque $\mathbb{W}$ in the complex quaternion space $\mathbb{H}_g$ , and the term $\mathbb{W}_e$ is the component of octonion torque $\mathbb{W}$ in the complex $S$-quaternion space $\mathbb{H}_e$ .

In the complex quaternion space $\mathbb{H}_g$ , the component $\mathbb{W}_g$ of octonion torque $\mathbb{W}$ is written as,
\begin{eqnarray}
\mathbb{W}_g = \mathbb{W}_g^g + k_{eg}^2 \mathbb{W}_e^e ~,
\end{eqnarray}
where $\mathbb{W}_g^g = - v_0 ( i \mathbb{F}_g / v_0 + \square_g ) \circ \mathbb{L}_g$ , $\mathbb{W}_e^e = - v_0 ( i \mathbb{F}_e / v_0 + k_{eg}^{-2} \square_e ) \circ \mathbb{L}_e$ .

The above can be rewritten as,
\begin{eqnarray}
\mathbb{W}_g = i W_{g0}^i + W_{g0} + i \textbf{W}_g^i + \textbf{W}_g ~,
\end{eqnarray}
and
\begin{eqnarray}
W_{g0}^i = && ( \textbf{g} \cdot \textbf{L}_g^i / v_0 - \textbf{b} \cdot \textbf{L}_g ) - v_0 ( \partial_{r0} L_{g0} + \nabla_r \cdot \textbf{L}_g^i)
\nonumber
\\
&&
+ k_{eg}^2 ( \textbf{E} \cdot \textbf{L}_e^i / v_0 - \textbf{B} \cdot \textbf{L}_e )
- v_0 ( \emph{\textbf{I}}_0 \partial_{R0} \circ \textbf{L}_{e0} + \nabla_R \cdot \textbf{L}_e^i ) ~,
\\
W_{g0} = && ( \textbf{b} \cdot \textbf{L}_g^i + \textbf{g} \cdot \textbf{L}_g / v_0 ) - v_0 ( \nabla_r \cdot \textbf{L}_g )
\nonumber
\\
&&
+ k_{eg}^2 ( \textbf{B} \cdot \textbf{L}_e^i + \textbf{E} \cdot \textbf{L}_e / v_0 ) - v_0 ( \nabla_R \cdot \textbf{L}_e ) ~,
\\
\textbf{W}_g^i = && ( \textbf{g} \times \textbf{L}_g^i / v_0 - L_{g0} \textbf{b} - \textbf{b} \times \textbf{L}_g )
- v_0 ( \partial_{r0} \textbf{L}_g + \nabla_r \times \textbf{L}_g^i )
\nonumber
\\
&&
+ k_{eg}^2 ( \textbf{E} \times \textbf{L}_e^i / v_0 - \textbf{B} \circ \textbf{L}_{e0} - \textbf{B} \times \textbf{L}_e )
\nonumber
\\
&&
- v_0 ( \emph{\textbf{I}}_0 \partial_{R0} \circ \textbf{L}_e + \nabla_R \times \textbf{L}_e^i ) ~,
\\
\textbf{W}_g = && ( \textbf{g} L_{g0} / v_0 + \textbf{g} \times \textbf{L}_g / v_0 + \textbf{b} \times \textbf{L}_g^i )
\nonumber
\\
&&
+ v_0 ( \partial_{r0} \textbf{L}_g^i - \nabla_r L_{g0} - \nabla_r \times \textbf{L}_g )
\nonumber
\\
&&
+ k_{eg}^2 ( \textbf{E} \circ \textbf{L}_{e0} / v_0 + \textbf{E} \times \textbf{L}_e / v_0 + \textbf{B} \times \textbf{L}_e^i )
\nonumber
\\
&&
+ v_0 ( \emph{\textbf{I}}_0 \partial_{R0} \circ \textbf{L}_e^i - \nabla_R \circ \textbf{L}_{e0} - \nabla_R \times \textbf{L}_e ) ~,
\end{eqnarray}
where $\textbf{W}_g = \Sigma W_{gk} \emph{\textbf{i}}_k$ . $\textbf{W}_g^i = \Sigma W^i_{gk} \emph{\textbf{i}}_k$ . $W_{g0}^i$ is the energy. $\textbf{W}_g^i$ is the torque. $\textbf{W}_g$ is the curl of angular momentum. $\{ k_{eg}^2 ( \textbf{E} \cdot \textbf{L}_e^i / v_0 - \textbf{B} \cdot \textbf{L}_e ) - v_0 ( \emph{\textbf{I}}_0 \partial_{R0} \circ \textbf{L}_{e0} + \nabla_R \cdot \textbf{L}_e^i ) \}$ is the contribution of the E adjoint-field to the energy. $\{ k_{eg}^2 ( \textbf{E} \times \textbf{L}_e^i / v_0 - \textbf{B} \circ \textbf{L}_{e0} - \textbf{B} \times \textbf{L}_e ) - v_0 ( \emph{\textbf{I}}_0 \partial_{R0} \circ \textbf{L}_e + \nabla_R \times \textbf{L}_e^i ) \}$ is the contribution of the E adjoint-field to the torque. $k_p = (k - 1)$ is a coefficient, and $k$ is the dimension of the vector $\textbf{r}$ . For $k_p = 2$, the term $( W_{g0}^i / 2 )$ is the energy in the three-dimensional space. Comparing this term $( W_{g0}^i / 2 )$ with the conventional energy in the classical field theory states, $k_{rx} = 1 / v_0$. $W_{gj}$ and $W^i_{gj}$ are all real.

\begin{table}[h]
\tbl{The physics quantities and definitions of the fundamental field and adjoint-field in the complex octonion space.}
{\begin{tabular}{@{}ll@{}}
\hline
octonion~physics~quantity ~~~~~~~~    &   definition                                                                                 \\
\hline
radius~vector                         &  $\mathbb{R} = \mathbb{R}_g + k_{eg} \mathbb{R}_e  $                                         \\
octonion~operator                     &  $\square = \square_g + k_{eg}^{-1} \square_e$                                               \\
integrating~function                  &  $\mathbb{X} = \mathbb{X}_g + k_{eg} \mathbb{X}_e  $                                         \\
field~potential                       &  $\mathbb{A} = i \square^\times \circ \mathbb{X}  $                                          \\
field~strength                        &  $\mathbb{F} = \square \circ \mathbb{A}  $                                                   \\
field~source                          &  $\mu \mathbb{S} = - ( i \mathbb{F} / v_0 + \square )^* \circ \mathbb{F} $                   \\
velocity operator                     &  $\square_V = \partial_{r0} + k_{eg}^{-1} \emph{\textbf{I}}_0 \partial_{R0}    $             \\
octonion velocity	                  &  $\mathbb{V} = v_0 \square_V \circ \mathbb{R}   $                                            \\
linear~momentum                       &  $\mathbb{P} = \mu \mathbb{S} / \mu_g^g $                                                    \\
angular~momentum                      &  $\mathbb{L} = ( \mathbb{R} + k_{rx} \mathbb{X} )^\times \circ \mathbb{P} $                  \\
octonion~torque                       &  $\mathbb{W} = - v_0 ( i \mathbb{F} / v_0 + \square ) \circ \mathbb{L} $                     \\
octonion~force                        &  $\mathbb{N} = - ( i \mathbb{F} / v_0 + \square ) \circ \mathbb{W} $                         \\
\hline
\end{tabular}}
\end{table}

In the complex $S$-quaternion space $\mathbb{H}_e$ , the component $\mathbb{W}_e$ is written as,
\begin{eqnarray}
\mathbb{W}_e = \mathbb{W}_e^g + \mathbb{W}_g^e ~,
\end{eqnarray}
where $\mathbb{W}_e^g = - v_0 ( i \mathbb{F}_g / v_0 + \square_g ) \circ \mathbb{L}_e$ , $\mathbb{W}_g^e = - v_0 ( i \mathbb{F}_e / v_0 + k_{eg}^{-2} \square_e ) \circ \mathbb{L}_g$ .

The above can be rewritten as,
\begin{eqnarray}
\mathbb{W}_e = i \textbf{W}_{e0}^i + \textbf{W}_{e0} + i \textbf{W}_e^i + \textbf{W}_e ~,
\end{eqnarray}
and
\begin{eqnarray}
\textbf{W}_{e0}^i = && ( \textbf{g} \cdot \textbf{L}_e^i / v_0 - \textbf{b} \cdot \textbf{L}_e )
- v_0 ( \partial_{r0} \textbf{L}_{e0} + \nabla_r \cdot \textbf{L}_e^i )
\nonumber
\\
&&
+ ( \textbf{E} \cdot \textbf{L}_g^i / v_0 - \textbf{B} \cdot \textbf{L}_g )
- v_0 k_{eg}^{-2} ( \emph{\textbf{I}}_0 \partial_{R0} L_{g0} + \nabla_R \cdot \textbf{L}_g^i ) ~,
\\
\textbf{W}_{e0} = && ( \textbf{b} \cdot \textbf{L}_e^i + \textbf{g} \cdot \textbf{L}_e / v_0 ) - v_0 ( \nabla_r \cdot \textbf{L}_e )
\nonumber
\\
&&
+ ( \textbf{B} \cdot \textbf{L}_g^i + \textbf{E} \cdot \textbf{L}_g / v_0 ) - v_0 k_{eg}^{-2} ( \nabla_R \cdot \textbf{L}_g ) ~,
\\
\textbf{W}_e^i = && ( \textbf{g} \times \textbf{L}_e^i / v_0 - \textbf{b} \circ \textbf{L}_{e0} - \textbf{b} \times \textbf{L}_e )
- v_0 ( \partial_{r0} \textbf{L}_e + \nabla_r \times \textbf{L}_e^i )
\nonumber
\\
&&
+ ( \textbf{E} \times \textbf{L}_g^i / v_0 - L_{g0} \textbf{B} - \textbf{B} \times \textbf{L}_g )
\nonumber
\\
&&
- v_0 k_{eg}^{-2} ( \emph{\textbf{I}}_0 \partial_{R0} \circ \textbf{L}_g + \nabla_R \times \textbf{L}_g^i ) ~,
\\
\textbf{W}_e = && ( \textbf{g} \circ \textbf{L}_{e0} / v_0 + \textbf{g} \times \textbf{L}_e / v_0 + \textbf{b} \times \textbf{L}_e^i )
\nonumber
\\
&&
+ v_0 ( \partial_{r0} \textbf{L}_e^i - \nabla_r \circ \textbf{L}_{e0} - \nabla_r \times \textbf{L}_e )
\nonumber
\\
&&
+ ( L_{g0} \textbf{E} / v_0 + \textbf{E} \times \textbf{L}_g / v_0 + \textbf{B} \times \textbf{L}_g^i )
\nonumber
\\
&&
+ v_0 k_{eg}^{-2} ( \emph{\textbf{I}}_0 \partial_{R0} \circ \textbf{L}_g^i - \nabla_R L_{g0} - \nabla_R \times \textbf{L}_g ) ~,
\end{eqnarray}
where $\textbf{W}_{e0}^i = W_{e0}^i \emph{\textbf{I}}_0$ . $\textbf{W}_{e0} = W_{e0} \emph{\textbf{I}}_0$ . $\textbf{W}_e = \Sigma W_{ek} \emph{\textbf{I}}_k$ . $\textbf{W}_{ei} = \Sigma W^i_{ek} \emph{\textbf{I}}_k$ . $\textbf{W}_{e0}$ is the divergence of magnetic dipole moment, and $\textbf{W}_e$ is the curl of magnetic dipole moment. $\{ ( \textbf{B} \cdot \textbf{L}_g^i + \textbf{E} \cdot \textbf{L}_g / v_0 ) - v_0 k_{eg}^{-2} ( \nabla_R \cdot \textbf{L}_g ) \}$ is the contribution of the G adjoint-field to the term $\textbf{W}_{e0}$ . $\{ ( L_{g0} \textbf{E} / v_0 + \textbf{E} \times \textbf{L}_g / v_0 + \textbf{B} \times \textbf{L}_g^i ) - v_0 k_{eg}^{-2} ( - \emph{\textbf{I}}_0 \partial_{R0} \circ \textbf{L}_g^i + \nabla_R L_{g0} + \nabla_R \times \textbf{L}_g ) \}$ is the contribution of the G adjoint-field to the term $\textbf{W}_e$ . $W_{ej}$ and $W^i_{ej}$ are all real.

In the electromagnetic and gravitational theories described with the complex octonion, the above states that the torque, energy, work, and the derivative of electric/magnetic dipole moment, are actually consistent in essence. The E adjoint-field (dark matter field) makes a contribution towards the energy, torque, and the curl of angular momentum and so on. And the G adjoint-field has the contribution to the divergence of and the curl of magnetic dipole moment and so forth. Obviously two adjoint-fields are in substantial agreement. The contrast study on these two adjoint-fields will be beneficial to improve further the comprehension to the property of E adjoint-field.

\begin{table}[h]
\tbl{The physics quantity of the electromagnetic field, gravitational field, and correlative adjoint-field in the complex octonion space.}
{\begin{tabular}{@{}ll@{}}
\hline
octonion~physics~quantity ~~~~~~~~    &  expression                                                                                                      \\
\hline
integrating~function	              &  $\mathbb{X} = \mathbb{X}_g + k_{eg} \mathbb{X}_e$ ~,                                                            \\
            	                      &  $\mathbb{X}_g = i x_0 + \Sigma \emph{\textbf{i}}_k x_k$ ~,~~
                                         $\mathbb{X}_e = i \emph{\textbf{I}}_0 X_0 + \Sigma \emph{\textbf{I}}_k X_k$ ;                                   \\
field~potential	                      &  $\mathbb{A} = \mathbb{A}_g + k_{eg} \mathbb{A}_e$ ~,                                                            \\
                	                  &  $\mathbb{A}_g = \mathbb{A}_g^g + \mathbb{A}_e^e$ ~,~~
                                         $\mathbb{A}_e = \mathbb{A}_e^g + k_{eg}^{-2} \mathbb{A}_g^e$ ;                                                  \\
field~strength	                      &  $\mathbb{F} = \mathbb{F}_g + k_{eg} \mathbb{F}_e$ ~,                                                            \\
	                                  &  $\mathbb{F}_g = \mathbb{F}_g^g + \mathbb{F}_e^e$ ~,~~
                                         $\mathbb{F}_e = \mathbb{F}_e^g + k_{eg}^{-2} \mathbb{F}_g^e$ ;                                                  \\
field~source	                      &  $\mu \mathbb{S} = \mu_g \mathbb{S}_g + k_{eg} \mu_e \mathbb{S}_e - i \mathbb{F}^* \circ \mathbb{F} / v_0 $ ~,   \\
	                                  &  $\mu_g \mathbb{S}_g = \mu_g^g \mathbb{S}_g^g + \mu_e^e \mathbb{S}_e^e$ ~,~~
                                         $\mu_e \mathbb{S}_e = \mu_e^g \mathbb{S}_e^g + k_{eg}^{-2} \mu_g^e \mathbb{S}_g^e$ ;                            \\
octonion~velocity	                  &  $\mathbb{V} = \mathbb{V}_g + k_{eg} \mathbb{V}_e$ ~,                                                            \\
	                                  &  $\mathbb{V}_g = \mathbb{V}_g^g + \mathbb{V}_e^e$ ~,~~
                                         $\mathbb{V}_e = \mathbb{V}_e^g + k_{eg}^{-2} \mathbb{V}_g^e$ ;                                                  \\
linear~momentum	                      &  $\mathbb{P} = \mathbb{P}_g + k_{eg} \mathbb{P}_e$ ~,                                                            \\
	                                  &  $\mathbb{P}_g = \mathbb{P}_g^g + \mathbb{P}_e^e$ ~,~~
                                         $\mathbb{P}_e = \mathbb{P}_e^g + k_{eg}^{-2} \mathbb{P}_g^e$ ;                                                  \\
angular~momentum	                  &  $\mathbb{L} = \mathbb{L}_g + k_{eg} \mathbb{L}_e$ ~,                                                            \\
	                                  &  $\mathbb{L}_g = \mathbb{L}_g^g + k_{eg}^2 \mathbb{L}_e^e$ ~,~~
                                         $\mathbb{L}_e = \mathbb{L}_e^g + \mathbb{L}_g^e$ ;                                                              \\
octonion~torque	                      &  $\mathbb{W} = \mathbb{W}_g + k_{eg} \mathbb{W}_e$ ~,                                                            \\
	                                  &  $\mathbb{W}_g = \mathbb{W}_g^g + k_{eg}^2 \mathbb{W}_e^e$ ~,~~
                                         $\mathbb{W}_e = \mathbb{W}_e^g + \mathbb{W}_g^e$ ;                                                              \\
octonion~force	                      &  $\mathbb{N} = \mathbb{N}_g + k_{eg} \mathbb{N}_e$ ~,                                                            \\
	                                  &  $\mathbb{N}_g = \mathbb{N}_g^g + k_{eg}^2 \mathbb{N}_e^e$ ~,~~
                                         $\mathbb{N}_e = \mathbb{N}_e^g + \mathbb{N}_g^e$ ;                                                              \\
\hline
\end{tabular}}
\end{table}

\section{Octonion force}	

In the electromagnetic and gravitational theories described with the complex octonion, the definition of octonion force covers the force and power in the complex quaternion space $\mathbb{H}_g$ , and the current continuity equation in the complex $S$-quaternion space $\mathbb{H}_e$ .

In the complex octonion space $\mathbb{O}$ , from the octonion operator $\square$ , octonion field strength $\mathbb{F}$ , and octonion torque $\mathbb{W}$ , the octonion force can be defined as,
\begin{eqnarray}
\mathbb{N} = - ( i \mathbb{F} / v_0 + \square ) \circ \mathbb{W} ~,
\end{eqnarray}
where $\mathbb{N} = \mathbb{N}_g + k_{eg} \mathbb{N}_e$ . The term $\mathbb{N}_g$ is the component of octonion force $\mathbb{N}$ in the complex quaternion space $\mathbb{H}_g$ , meanwhile the term $\mathbb{N}_e$ is the component of octonion force $\mathbb{N}$ in the complex $S$-quaternion space $\mathbb{H}_e$ .

In the complex quaternion space $\mathbb{H}_g$ , the component $\mathbb{N}_g$ of the octonion force is written as,
\begin{eqnarray}
\mathbb{N}_g = \mathbb{N}_g^g + k_{eg}^2 \mathbb{N}_e^e ~,
\end{eqnarray}
where $\mathbb{N}_g^g = - ( i \mathbb{F}_g / v_0 + \square_g ) \circ \mathbb{W}_g$ , $\mathbb{N}_e^e = - ( i \mathbb{F}_e / v_0 + k_{eg}^{-2} \square_e ) \circ \mathbb{W}_e$ .

The above can be written as,
\begin{eqnarray}
\mathbb{N}_g = i N_{g0}^i + N_{g0} + i \textbf{N}_g^i + \textbf{N}_g ~,
\end{eqnarray}
and
\begin{eqnarray}
N_{g0}^i = && ( \textbf{g} \cdot \textbf{W}_g^i / v_0 - \textbf{b} \cdot \textbf{W}_g ) / v_0
- ( \partial_{0r} W_{g0} + \nabla_r \textbf{W}_g^i )
\nonumber
\\
&&
+ k_{eg}^2 ( \textbf{E} \cdot \textbf{W}_e^i / v_0 - \textbf{B} \cdot \textbf{W}_e ) / v_0
- ( \emph{\textbf{I}}_0 \partial_{R0} \circ \textbf{W}_{e0} + \nabla_R \cdot \textbf{W}_e^i ) ~,
\\
N_{g0} = && ( \textbf{g} \cdot \textbf{W}_g / v_0 + \textbf{b} \cdot \textbf{W}_g^i ) / v_0
+ ( \partial_{0r} W_{g0}^i - \nabla_r \cdot \textbf{W}_g )
\nonumber
\\
&&
+ k_{eg}^2 ( \textbf{E} \cdot \textbf{W}_e / v_0 + \textbf{B} \cdot \textbf{W}_e^i ) / v_0 - ( \nabla_R \cdot \textbf{W}_e ) ~,
\\
\textbf{N}_g^i = && ( W_{g0}^i \textbf{g} / v_0 + \textbf{g} \times \textbf{W}_g^i / v_0 - W_{g0} \textbf{b} - \textbf{b} \times \textbf{W}_g ) / v_0
\nonumber
\\
&&
- ( \partial_{0r} \textbf{W}_g + \nabla_r W_{g0}^i + \nabla_r \times \textbf{W}_g^i )
- ( \emph{\textbf{I}}_0 \partial_{R0} \circ \textbf{W}_e + \nabla_R \times \textbf{W}_e^i )
\nonumber
\\
&&
+ k_{eg}^2 ( \textbf{E} \circ \textbf{W}_{e0}^i / v_0 + \textbf{E} \times \textbf{W}_e^i / v_0 - \textbf{B} \circ \textbf{W}_{e0}
- \textbf{B} \times \textbf{W}_e ) / v_0 ~,
\\
\textbf{N}_g = && ( W_{g0} \textbf{g} / v_0 + \textbf{g} \times \textbf{W}_g / v_0 + W_{g0}^i \textbf{b} + \textbf{b} \times \textbf{W}_g^i ) / v_0
\nonumber
\\
&&
+ ( \partial_{0r} \textbf{W}_g^i - \nabla_r W_{g0} - \nabla_r \times \textbf{W}_g )
\nonumber
\\
&&
+ k_{eg}^2 ( \textbf{E} \circ \textbf{W}_{e0} / v_0 + \textbf{E} \times \textbf{W}_e/ v_0 + \textbf{B} \circ \textbf{W}_{e0}^i + \textbf{B} \times \textbf{W}_e^i ) / v_0
\nonumber
\\
&&
+ ( \emph{\textbf{I}}_0 \partial_{R0} \circ \textbf{W}_e^i - \nabla_R \circ \textbf{W}_{e0} - \nabla_R \times \textbf{W}_e ) ~,
\end{eqnarray}
where $\textbf{N}_g = \Sigma N_{gk} \emph{\textbf{i}}_k$ . $\textbf{N}_g^i = \Sigma N^i_{gk} \emph{\textbf{i}}_k$ . $N_{g0}$ is the power, $\textbf{N}_g^i$ is the force, and $\textbf{N}_g$ is the derivative of torque. Comparing the term $( \textbf{N}_g^i / k_p )$ with the force in the classical field theory states, $k_{eg}^2 = \mu_g^g / \mu_e^g < 0$. $\{ k_{eg}^2 ( \textbf{E} \cdot \textbf{W}_e / v_0 + \textbf{B} \cdot \textbf{W}_e^i ) / v_0 - ( \nabla_R \cdot \textbf{W}_e ) \}$ is the contribution of the E adjoint-field to the power. $\{ k_{eg}^2 ( \textbf{E} \circ \textbf{W}_{e0}^i / v_0 + \textbf{E} \times \textbf{W}_e^i / v_0 - \textbf{B} \circ \textbf{W}_{e0} - \textbf{B} \times \textbf{W}_e ) / v_0 - ( \emph{\textbf{I}}_0 \partial_{R0} \circ \textbf{W}_e + \nabla_R \times \textbf{W}_e^i ) \}$ is the contribution of the E adjoint-field to the force. $\{  k_{eg}^2 ( \textbf{E} \circ \textbf{W}_{e0} / v_0 + \textbf{E} \times \textbf{W}_e/ v_0 + \textbf{B} \circ \textbf{W}_{e0}^i + \textbf{B} \times \textbf{W}_e^i ) / v_0 - ( - \emph{\textbf{I}}_0 \partial_{R0} \circ \textbf{W}_e^i + \nabla_R \circ \textbf{W}_{e0} + \nabla_R \times \textbf{W}_e )  \}$ is the contribution of the E adjoint-field to the derivative of torque. $N_{gj}$ and $N^i_{gj}$ are all real.

In the above, the expression of force, $( \textbf{N}_g^i / k_p )$, reveals that the E adjoint-field source (dark matter particle) can be influenced by the gravitational strength, and the G adjoint-field source can be impacted by the electromagnetic strength. Moreover, the part of energy gradient, which is relevant to the E adjoint-field, is also one component of the force.

In the complex $S$-quaternion space $\mathbb{H}_e$ , the component $\mathbb{N}_e$ is written as,
\begin{eqnarray}
\mathbb{N}_e = \mathbb{N}_e^g + \mathbb{N}_g^e ~,
\end{eqnarray}
where $\mathbb{N}_e^g = - ( i \mathbb{F}_g / v_0 + \square_g ) \circ \mathbb{W}_e$ , $\mathbb{N}_g^e = - ( i \mathbb{F}_e / v_0 + k_{eg}^{-2} \square_e ) \circ \mathbb{W}_g$ .

The above will be rewritten as,
\begin{eqnarray}
\textbf{N}_e = i \textbf{N}_{e0}^i + \textbf{N}_{e0} + i \textbf{N}_e^i + \textbf{N}_e ~,
\end{eqnarray}
and
\begin{eqnarray}
\textbf{N}_{e0}^i = && ( \textbf{g} \cdot \textbf{W}_e^i / v_0 - \textbf{b} \cdot \textbf{W}_e ) / v_0
- ( \partial_{0r} \textbf{W}_{e0} + \nabla_r \cdot \textbf{W}_e^i )
\nonumber
\\
&&
+ ( \textbf{E} \cdot \textbf{W}_g^i / v_0 - \textbf{B} \cdot \textbf{W}_g ) / v_0
- k_{eg}^{-2} ( \emph{\textbf{I}}_0 \partial_{R0} W_{g0} + \nabla_R \cdot \textbf{W}_g^i ) ~,
\\
\textbf{N}_{e0} = && ( \textbf{g} \cdot \textbf{W}_e / v_0 + \textbf{b} \cdot \textbf{W}_e^i ) / v_0
+ ( \partial_{0r} \textbf{W}_{e0}^i - \nabla_r \cdot \textbf{W}_e )
\nonumber
\\
&&
+ ( \textbf{E} \cdot \textbf{W}_g / v_0 + \textbf{B} \cdot \textbf{W}_g^i ) / v_0 - k_{eg}^{-2} ( \nabla_R \cdot \textbf{W}_g ) ~,
\\
\textbf{N}_e^i = && ( \textbf{g} \circ \textbf{W}_{e0}^i / v_0 + \textbf{g} \times \textbf{W}_e^i / v_0 - \textbf{b} \circ \textbf{W}_{e0} - \textbf{b} \times \textbf{W}_e ) / v_0
\nonumber
\\
&&
- ( \partial_{0r} \textbf{W}_e +  \nabla_r \circ \textbf{W}_{e0}^i + \nabla_r \times \textbf{W}_e^i )
\nonumber
\\
&&
+ ( W_{g0}^i \textbf{E} / v_0 + \textbf{E} \times \textbf{W}_g^i / v_0 - W_{g0} \textbf{B} - \textbf{B} \times \textbf{W}_g ) / v_0
\nonumber
\\
&&
- k_{eg}^{-2} ( \emph{\textbf{I}}_0 \partial_{R0} \circ \textbf{W}_g + \nabla_R \times \textbf{W}_g^i )  ~,
\\
\textbf{N}_e = && ( \textbf{g} \circ \textbf{W}_{e0} / v_0 + \textbf{g} \times \textbf{W}_e / v_0 + \textbf{b} \circ \textbf{W}_{e0}^i + \textbf{b} \times \textbf{W}_e^i ) / v_0
\nonumber
\\
&&
+ ( \partial_{0r} \textbf{W}_e^i - \nabla_r \circ \textbf{W}_{e0} - \nabla_r \times \textbf{W}_e )
\nonumber
\\
&&
+ ( W_{g0} \textbf{E} / v_0 + \textbf{E} \times \textbf{W}_g / v_0 + W_{g0}^i \textbf{B} + \textbf{B} \times \textbf{W}_g^i ) / v_0
\nonumber
\\
&&
+ k_{eg}^{-2} ( \emph{\textbf{I}}_0 \partial_{R0} \circ \textbf{W}_g^i - \nabla_R W_{g0} - \nabla_R \times \textbf{W}_g ) ~,
\end{eqnarray}
where $\textbf{N}_{e0}^i = N_{e0}^i \emph{\textbf{I}}_0$ . $\textbf{N}_{e0} = N_{e0} \emph{\textbf{I}}_0$ . $\textbf{N}_e = \Sigma N_{ek} \emph{\textbf{I}}_k$ . $\textbf{N}_e^i = \Sigma N^i_{ek} \emph{\textbf{I}}_k$ . $N_{e0}$ is similar to the power. $\{ ( \textbf{E} \cdot \textbf{W}_g / v_0 + \textbf{B} \cdot \textbf{W}_g^i ) / v_0 - k_{eg}^{-2} ( \nabla_R \cdot \textbf{W}_g ) \}$ is the contribution of the G adjoint-field to the term $N_{e0}$ . $N_{ej}$ and $N^i_{ej}$ are all real.

Under certain conditions, there may be $\mathbb{N} = 0$, including $\textbf{N}_g^i = 0$, $\textbf{N}_g = 0$, $\textbf{N}_{g0} = 0$, and $\textbf{N}_{e0} = 0$ and so on. From $\textbf{N}_g^i = 0$, it may be able to deduce the force equilibrium equation. And the E adjoint-field makes a contribution to the force. From $\textbf{N}_g = 0$, it is capable of inferring the equation of precessional angular velocity. Also the E adjoint-field has one contribution to this equation. Meanwhile, it is able to figure out the mass (or linear momentum) continuity equation, from $\textbf{N}_{g0} = 0$. And it can conclude the current continuity equation, from $\textbf{N}_{e0} = 0$. Expressly, the adjoint-field makes a contribution to both of two continuity equations.

When both of G and E adjoint-fields can be neglected, the field strength, field source, linear momentum, angular momentum, torque, and force and so forth in the above, can be degenerated respectively into that of the field theory without the operator $\square_e$ , in the complex octonion space. And these physics quantities will be simplified sequentially into that in the classical field theory, when the adjoint-fields are quite weak.

\section{Dark matter property}

If the E adjoint-field is the dark matter field, the latter will become an important and indispensable component of the electromagnetic and gravitational theories in the complex octonion space. As a result, it is able to predict theoretically a few deductions for the properties of dark matter, according to this field theory described with the complex octonion.

In the complex octonion space, by means of the $S$-quaternion operator, it is capable of producing two adjoint-fields, the E adjoint-field and G adjoint-field. However, these two adjoint-fields can not be self-existent, according to the octonion multiplication. When the $S$-quaternion operator acts on the physics quantities of adjoint-fields, these physics quantities will be transferred alternatively between the quaternion space and $S$-quaternion space. These two adjoint-fields are interlocking and dependent on each other, and have one kind of coexistent relationship.

The E adjoint-field possesses its own field source and particle. And these physics quantities are independent of that of gravitational field, electromagnetic field, or G adjoint-field. To a certain extent, the E adjoint-field may make a contribution to the complex octonion linear momentum, angular momentum, torque, energy, force, and power and so on. Of course, the E adjoint-field can combine with the gravitational field, exerting an influence on the movements and diversification of the ordinary matter correspondingly (Table 6).

\begin{table}[h]
\tbl{The physics quantity and definition of fundamental fields in the complex octonion space.}
{\begin{tabular}{@{}lll@{}}
\hline
physics~quantity	&  gravitational~field                                                                    &  electromagnetic~field                                                               \\
                    &  (quaternion~space)                                                                     &  ($S$-quaternion~space)                                                              \\
\hline
field~potential	    &  $\mathbb{A}_g^g = i \square_g^\times \circ \mathbb{X}_g$	                              &  $\mathbb{A}_e^g = i \square_g^\times \circ \mathbb{X}_e$                            \\
field~strength	    &  $\mathbb{F}_g^g = \square_g \circ \mathbb{A}_g$	                                      &  $\mathbb{F}_e^g = \square_g \circ \mathbb{A}_e$                                     \\
field~equations	    &  $\mu_g^g \mathbb{S}_g^g = - \square_g^* \circ \mathbb{F}_g$	                          &  $\mu_e^g \mathbb{S}_e^g = - \square_g^* \circ \mathbb{F}_e$                         \\
field~source	    &  $\mathbb{S}_g^g = m_g^g \mathbb{V}_g^g$	                                              &  $\mathbb{S}_e^g = m_e^g \mathbb{V}_e^g$                                             \\
octonion~velocity	&  $\mathbb{V}_g^g = \partial_{r0} \mathbb{R}_g$	                                      &  $\mathbb{V}_e^g = \partial_{r0} \mathbb{R}_e$                                       \\
linear~momentum	    &  $\mathbb{P}_g^g = \mathbb{S}_g^g - i \mathbb{F}^* \circ \mathbb{F} / ( \mu_g^g v_0 )$  &  $\mathbb{P}_e^g = \mu_e^g \mathbb{S}_e^g / \mu_g^g$                                 \\
angular~momentum	&  $\mathbb{L}_g^g = ( \mathbb{R}_g + k_{rx} \mathbb{X}_g )^\times \circ \mathbb{P}_g$    &  $\mathbb{L}_e^g = ( \mathbb{R}_g + k_{rx} \mathbb{X}_g )^\times \circ \mathbb{P}_e$ \\
octonion~torque	    &  $\mathbb{W}_g^g = - v_0 ( i \mathbb{F}_g / v_0 + \square_g ) \circ \mathbb{L}_g$	      &  $\mathbb{W}_e^g = - v_0 ( i \mathbb{F}_g / v_0 + \square_g ) \circ \mathbb{L}_e$    \\
octonion~force	    &  $\mathbb{N}_g^g = - ( i \mathbb{F}_g / v_0 + \square_g ) \circ \mathbb{W}_g$	          &  $\mathbb{N}_e^g = - ( i \mathbb{F}_g / v_0 + \square_g ) \circ \mathbb{W}_e$        \\
\hline
\end{tabular}}
\end{table}

\subsection{Weak intensity}

According to existing studies, it is reasonable to suppose that the interaction intensity of dark matter field (E adjoint-field) is really weak, and is weaker than that of gravitational field. But it may not be at the end of the queue, and could be a little stronger than that of G adjoint-field. Otherwise, the interaction intensity of dark matter field would be strong enough, to result in a large block of dark matters with comparatively high density. But it defies the astronomic observations. So the interaction coefficients may be,
\begin{eqnarray}
\mid \mu_e^g  \mid  > \mid \mu_g^g  \mid  >  \mid  \mu_e^e  \mid   >   \mid  \mu_g^e  \mid ~.
\end{eqnarray}

In the electromagnetic and gravitational theories described with the complex octonion, the gravitational coefficient is, $\mu_g^g < 0$, and the electromagnetic coefficient is, $\mu_e^g > 0$. For the adjoint-field of a fundamental field, there is one kind of possibility: the E adjoint-field coefficient may be, $\mu_e^e < 0$, and the G adjoint-field coefficient may be, $\mu_g^e > 0$. Subsequently the emergence of E adjoint-field will amplify the interaction intensity of gravitational field, increasing properly the energy, torque, and angular momentum and so forth of the galaxy.

Therefore the gravitational coefficient $\mu_g^g$ and the E adjoint-field coefficient $\mu_e^e$ both have an attracting effect, under the stable state of movement. And it enables some rotational galaxies to be acquired one extra part of centripetal force.

\subsection{One-source particle}

In the complex octonion space $\mathbb{O}$ , there are four kinds of `charges', including the `charge' (mass, $m_g^g$) of gravitational field, the `charge' (electric charge, $m_e^g$) of electromagnetic field, the `charge' ($m_g^e$) of G adjoint-field, and the `charge' ($m_e^e$) of E adjoint-field. For lack of the effective measurement technology, a majority of the particles relevant to the `four charges' are unable to be measured at the moment. The measurement technology in the astronomy up to now, are mainly dependent on the measurement approach of photon (optics and radioastrophysics). Only the electric charge is correlative with the photon, among all of `four charges'. Consequently there is a great deal of particles relevant to the `four charges', which are unobservable temporarily.

Making use of the measurement of gravitational strength $\mathbb{F}_g^g$ , what one can make certain is just a part of the component $\mathbb{S}_g^g$ of gravitational source. Further it is necessary to measure the E adjoint-field strength $\mathbb{F}_e^e$ , in order to make sure the rest of component $\mathbb{S}_g^g$ . Simultaneously measuring these two field strengths, $\mathbb{F}_g^g$ and $\mathbb{F}_e^e$ , may ascertain the whole component $\mathbb{S}_g^g$ of the gravitational source. While the component $\mathbb{S}_g^g$ is merely a portion of the gravitational source $\mathbb{S}_g$ .

For the component $\mathbb{S}_e^g$ of electromagnetic source, the situation is as similar as possible to the above. By means of the measurement of electromagnetic strength $\mathbb{F}_e^g$ , what one may ensure is just a part of the component $\mathbb{S}_e^g$ of electromagnetic source. Further it is necessary to measure the G adjoint-field strength $\mathbb{F}_g^e$ , for the sake of confirming the rest of the component $\mathbb{S}_e^g$ of electromagnetic source. Measuring synchronously these two field strengths, $\mathbb{F}_g^e$ and $\mathbb{F}_e^g$ , may make clear the whole component $\mathbb{S}_e^g$ of the electromagnetic source. And the component $\mathbb{S}_e^g$ is only a portion of electromagnetic source $\mathbb{S}_e$ .

\begin{table}[h]
\tbl{The physics quantity and definition of adjoint-fields in the complex octonion space.}
{\begin{tabular}{@{}lll@{}}
\hline
physics~quantity	&  E~adjoint-field	                                                                      &  G~adjoint-field                                                                     \\
                    &  (quaternion~space)                                                                     &  ($S$-quaternion~space)                                                              \\
\hline
field~potential	    &  $\mathbb{A}_e^e = i \square_e^\times \circ \mathbb{X}_e$	                              &  $\mathbb{A}_g^e = i \square_e^\times \circ \mathbb{X}_g$                            \\
field~strength	    &  $\mathbb{F}_e^e = \square_e \circ \mathbb{A}_e$                                        &  $\mathbb{F}_g^e = \square_e \circ \mathbb{A}_g$                                     \\
field~equations	    &  $\mu_e^e \mathbb{S}_e^e = - \square_e^* \circ \mathbb{F}_e$                            &  $\mu_g^e \mathbb{S}_g^e = - \square_e^* \circ \mathbb{F}_g$                         \\
field~source        &  $\mathbb{S}_e^e = m_e^e \mathbb{V}_e^e$                                                &  $\mathbb{S}_g^e = m_g^e \mathbb{V}_g^e$                                             \\
octonion~velocity   &  $\mathbb{V}_e^e = \emph{\textbf{I}}_0 \partial_{R0} \circ \mathbb{R}_e$                &  $\mathbb{V}_g^e = \emph{\textbf{I}}_0 \partial_{R0} \circ \mathbb{R}_g$             \\
linear~momentum     &  $\mathbb{P}_e^e = \mu_e^e \mathbb{S}_e^e / \mu_g^g$                                    &  $\mathbb{P}_g^e = \mu_g^e \mathbb{S}_g^e / \mu_g^g$                                 \\
angular~momentum    &  $\mathbb{L}_e^e = ( \mathbb{R}_e + k_{rx} \mathbb{X}_e )^\times \circ \mathbb{P}_e$    &  $\mathbb{L}_g^e = ( \mathbb{R}_e + k_{rx} \mathbb{X}_e )^\times \circ \mathbb{P}_g$ \\
octonion~torque     &  $\mathbb{W}_e^e = - v_0 ( i \mathbb{F}_e / v_0 + k_{eg}^{-2} \square_e ) \circ \mathbb{L}_e$  &  $\mathbb{W}_g^e = - v_0 ( i \mathbb{F}_e / v_0 + k_{eg}^{-2} \square_e ) \circ \mathbb{L}_g$             \\
octonion~force      &  $\mathbb{N}_e^e = - ( i \mathbb{F}_e / v_0 + k_{eg}^{-2} \square_e ) \circ \mathbb{W}_e$      &  $\mathbb{N}_g^e = - ( i \mathbb{F}_e / v_0 + k_{eg}^{-2} \square_e ) \circ \mathbb{W}_g$             \\
\hline
\end{tabular}}
\end{table}

\subsection{Two-source particle}	

In the electromagnetic and gravitational theories described with complex octonion, for the two-source particles, there are six kinds of conceivable combinations with the `four charges', including ($m_g^g$ , $m_e^g$), ($m_g^g$ , $m_g^e$), ($m_g^g$ , $m_e^e$), ($m_e^g$ , $m_g^e$), ($m_e^g$ , $m_e^e$), and ($m_g^e$ , $m_e^e$). In the distant interstellar space, only the particles relevant to the electric charge ($m_e^g$) can be observed at the present time. When the electromagnetic source combines with the gravitational source to become one charged particle, the latter will be simultaneously influenced by both of electromagnetic and gravitational fields. For the familiar charged particle ($m_g^g$ , $m_e^g$), there are already several effective detection approaches for the observations in the astronomy.

For the two-source particles, three of them are associated with the E adjoint-field. (i) The first particle is one kind of `charged particle', which contains the electromagnetic source and E adjoint-field source. And it is impacted by not only the electromagnetic field but also the E adjoint-field. When the electromagnetic source combines with the E adjoint-field source to become one `charged particle', the latter will be influenced by both of electromagnetic field and E adjoint-field simultaneously. (ii) The second particle comprises the gravitational source and E adjoint-field source. And it will be attracted by both of gravitational field and E adjoint-field simultaneously. (iii) The third particle is quite daunting to be observed, and consists of the E adjoint-field source and G adjoint-field source. This kind of particle can be affected by neither the gravitational field nor the electromagnetic field, appealing to develop a few new observation approaches.

In the complex octonion space, the measurement of particles relevant to the electric charge, will encounter one special `three-color' problem. For the electric charge ($m_e^g$), there may be three kinds of two-source particles, including ($m_g^g$ , $m_e^g$), ($m_e^g$ , $m_g^e$), and ($m_e^g$ , $m_e^e$). In fact, it is similar to the `three-color' phenomenon in the Quark theory. For the rest of `four charges', each charge has the similar `three-color' problem also.

However only the electric charge ($m_e^g$) may emit the photon sometimes. The scholars are incapable of measuring the messages emitted from the rest of `four charges' temporarily. For these charges existed in the universe, there are not effective detection approaches, but this is not the half of it. So it is necessary to develop some new distant measurement technologies, exploring the undiscovered matters of the universe. Undoubtedly the existence of these undiscovered matters will play an important role in the study of the structure and evolution of the universe.

\subsection{Long-range field}

As the fundamental field, the gravitational field and electromagnetic field both belong to the long-range field. Naturally one may suppose that the adjoint-field of one fundamental field is one long-range field. In the field theory, each fundamental field possesses its own intermediate boson. In a similar way, it is able to reckon that each adjoint-field has its own intermediate boson, which is the particle with the zero rest mass and is not the photon. Therefore the E adjoint-field is independent of the photon (intermediate boson). And it neither emits nor absorbs the light or the electromagnetic wave. Similarly the G adjoint-field has also its own intermediate boson without the rest mass, and it is independent of the photon either.

The photon is one intermediate boson of the electromagnetic field. And it will be severely and closely influenced by the electromagnetic field. As one particle with the energy, the photon will also be affected by the gravitational field to a certain extent. Following the same idea, the E adjoint-field will impact the photon partly. Obviously its interaction intensity on the photon should be weaker than that of gravitational field, such as the gravitational lensing.

The force equilibrium equation reveals that the E adjoint-field source and gravitational source both will be influenced by the force and gravitational strength, so the existence of E adjoint-field source will impact distantly the movement state of ordinary matter in the universe. For the fundamental field, the adjoint-field is just a natural concomitant, and can not be self-existent. As the supplement to one fundamental field source, the E adjoint-field source makes a contribution to the linear momentum, angular momentum, energy, torque, and force and so on of the galaxy to a certain extent.

If the E adjoint-field is the dark matter field, it is one appropriate method to combine the dark matter field with strong gravitational field, to become one research direction. Similarly it may be one conceivable combination to mix the dark matter field and strong electromagnetic field, becoming one new research direction. In particular the ultrastrong E adjoint-field will exert an influence on the mass and linear momentum distinctly. And the superstrong G adjoint-field may have an influence on the electric charge and current evidently.

\subsection{Galactic rotation curve}

For each fundamental field (gravitational or electromagnetic field), the emergence of adjoint-field is accompanied by the existence of fundamental field. The occurrence of adjoint-field enhances the interaction intensity of one fundamental field, resulting in some anomalous movement phenomena. Such as, the appearance of E adjoint-field (dark matter) leads to a few anomalous rotation curves of some galaxies.

In the vicinity of the galactic nucleus, the existence of E adjoint-field (dark matter) increases the mass of galactic nucleus, amplifying the attraction on the stars near the galaxy edge. On the other hand, for the stars on the edge of the galaxy, the emergence of E adjoint-field around one star intensifies also the star's gravity exerted by the galactic nucleus, accelerating the rotation speed of these stars and even their nearby E adjoint-fields.

In the gravitational field of one galaxy, if there also is the E adjoint-field strength, it means that the electromagnetic potential $\mathbb{A}_e$ will play an important role in the galaxy, since the E adjoint-field strength and the electromagnetic strength both are derived from the electromagnetic potential $\mathbb{A}_e$ . In this case, the activity level of electromagnetic source of the galaxy may be awfully fierce. As a result, the influence of E adjoint-field strength of one galaxy will be prominent for some galaxies. And it can heighten the interaction intensity of gravity, inducing the anomalous phenomenon in the galactic rotation curve. Moreover the electromagnetic strength should be comparatively strong in these galaxies also.

However, for some galaxies, their influences of E adjoint-field strengths are comparatively weak, and their contributions of E adjoint-field strengths to the interaction intensity of gravity are tiny. Accordingly the galactic rotation curve is normal. And the electromagnetic strength of galaxy may be comparatively faint.

\begin{table}[h]
\tbl{Comparison of physics properties between the E adjoint-field with dark matter field.}
{\begin{tabular}{@{}lll@{}}
\hline
field	                      &  E~adjoint-field	                                                       &  dark matter field                                                    \\
\hline
common origin	              &  the emergence is accompanied by the 	                                   &  (not yet)                                                            \\
                              &  existence of electromagnetic field                                                                                                                \\
interaction intensity	      &  weak strength	                                                           &  weak strength                                                        \\
long-range field	          &  long-range field	                                                       &  long-range field                                                     \\
intermediate boson	          &  independent of the photon                                                 &  independent of the photon                                            \\
appearance	                  &  enhance the gravity	                                                   &  speed up the galaxy rotation                                         \\
particle category	          &  one kind of one-source particle, and                                      &  (not yet)                                                            \\
                              &  three kinds of two-source particles                                                                                                               \\
force property	              &  similar to gravity 	                                                   &  similar to gravity                                                   \\
existence region	          &  region with the ultrastrong field                                         &  galaxy and cluster                                                   \\
                              &  strength, including the white dwarf,                                                                                                              \\
                              &  neutral star (magnetic star and pulsar),                                                                                                          \\
                              &  galactic nucleus, black hole, and astro-                                                                                                          \\
                              &  physical jet, point charge, atomic                                                                                                                \\
                              &  nucleus, and strong magnetic field                                                                                                                \\
\hline
\end{tabular}}
\end{table}

\section{Experiment proposal}

According to the gravitational and electromagnetic theories described with the complex octonion, it is capable of speculating about some possible existence regions of the E adjoint-field source (dark matter field source) and of the G adjoint-field source. For the study of G adjoint-field source, it has to be observed distantly in the astronomy. But fortunately for the investigation of the E adjoint-field source, it is probable to develop the relevant researches in the laboratory, besides the observing distantly in the astronomy.

\subsection{E adjoint-field}

The E adjoint-field possesses a part of gravitational properties. In the laboratory, it may be able to explore it experimentally. Making use of the improvements of existing technologies, one can manipulate the variation of the electromagnetic strength, which is sufficient to alter the electromagnetic source (in the $S$-quaternion space), and even the E adjoint-field source (in the quaternion space).

In the laboratory, when the electric field intensity $\textbf{E}$ is strong enough, it is possible to shift the E adjoint-field source, in the region surrounding the point charge or the atomic nucleus. In the quaternion/$S$-quaternion space, by means of transferring and measuring the alternative frequency and the directional gradient of electromagnetic strength, one may transfer the E adjoint-field source. Similarly it is capable of altering the E adjoint-field source, when the magnetic flux density $\textbf{B}$ is strong enough, in the magnetic confinement fusion device.

In the astronomic observation, when the magnetic flux density $\textbf{B}$ is comparative strong and its variation is fierce enough, there may be the E adjoint-field source, in the region enclosing the white dwarf and the neutral star, especially the magnetic star and pulsar. In other words, there may be some extra gravitational phenomena nearby these heavenly bodies.

\subsection{G adjoint-field}

The G adjoint-field owns one part of electromagnetic properties. At the moment, it has to be observed distantly in the astronomy. The existing technology is insufficient to develop the relevant investigations for the G adjoint-field in the laboratory, since it is too difficult to vary the gravitational strength. And therefore it is unable to alter the gravitational source (in the quaternion space), nor the G adjoint-field source (in the $S$-quaternion space).

In the astronomic observation, when the gravitational acceleration $\textbf{g}$ (correspond to the linear acceleration) is comparative strong, there may be the G adjoint-field source, in the region encompassing the galactic nucleus and black hole. When the gravitational strength component $\textbf{b}$ (correspond to the precessional angular velocity) is strong enough, there may be the G adjoint-field source, in the region surrounding the astrophysics jet and so on. That is, there may be a few additional electromagnetic phenomena surrounding these celestial bodies.

\section{Conclusions and discussions}	

One octonion space can be separated into two orthogonal subspaces, the quaternion space and $S$-quaternion space. The application of the quaternion operator reveals that the quaternion space is proper to describe the physics property of gravitational field, while the $S$-quaternion space is fit for depicting the features of electromagnetic field. With the $S$-quaternion operator instead of the quaternion operator, it is able to produce two adjoint-fields, the G adjoint-field and E adjoint-field. In other words, the application of the octonion operator and complex octonion space will be capable of exploring the field potential, field strength, field source, linear momentum, angular momentum, dipole moment, torque, energy, and force and so forth, in the electromagnetic field, gravitational field, and two adjoint-fields.

The adjoint-field possesses a part of the physics property of one fundamental field. In particular the E adjoint-field owns part properties of gravitational field. So the E adjoint-field can be chosen as one candidate of the dark matter field. From the viewpoint of the integrating function of field potential, the dark matter (E adjoint-field) and the ordinary matter both have the common origin. In terms of the complex $S$-quaternion operator and octonion space, it is able to deduce the contribution of dark matters to some physics quantities, including the field strength, field source, angular momentum, torque, and force and so forth. Moreover one may inference the particle category of dark matters, according to the theoretical model for the one-source and two-source particles.

The present detection approaches for the dark matter are mainly divided into two classes, the astronomic observation and the experimental exploration. (i) In the astronomic observation, the dark matter may exist in some regions nearby the celestial bodies with the ultrastrong field strength, including the white dwarf, neutral star, black hole, and astrophysical jet and so on. When the gravitational acceleration $\textbf{g}$ is comparative strong, it ought to observe the region close to the galactic nucleus and black hole, to find the G adjoint-field source. When the gravitational strength component $\textbf{b}$ is strong enough, it should be observed the region surrounding the astrophysics jet, to explore the G adjoint-field source. When the magnetic flux density $\textbf{B}$ is quite strong, it will be observed the region enclosing the white dwarf and the neutral star, especially the magnetic star and pulsar, to seek the E adjoint-field source. (ii) In the experimental exploration, making use of transforming the electromagnetic strength, it may be able to detect the E adjoint-field source. When the electric field intensity $\textbf{E}$ is strong enough, it could be investigated the region surrounding the point charge and atomic nucleus, to measure the E adjoint-field source. When the magnetic flux density $\textbf{B}$ is strong enough, one can observe the region surrounding the ultrastrong magnetic field inside the magnetic confinement fusion device, to probe the E adjoint-field source.

In the complex octonion space, a majority of physics quantities are able to be measured, including the force and linear acceleration and so on in the complex quaternion space, and the electromagnetic strength and electromagnetic source and so forth in the complex $S$-quaternion space. In the complex quaternion space, the physics quantities in the gravitational field are able to be measured directly, including the radius vector $\mathbb{R}_g$ , velocity $\mathbb{V}_g$ , linear acceleration, angular velocity, gravitational acceleration, linear momentum $\mathbb{P}_g$ , work, force, torque, power, and angular momentum $\mathbb{L}_g$ and so forth. In the complex $S$-quaternion space, the most of physics quantities relevant to the electromagnetic field may only be detected indirectly, including the electric field intensity, magnetic flux density, electric/magnetic dipole moment, electric charge, and electric current and so forth. By means of the force equilibrium equation, it is able to determine the Lorentz force. Similarly it is capable of calculating the radius vector $\mathbb{R}_e$ , and the velocity $\mathbb{V}_e$ in the complex $S$-quaternion space, making use of the measurement of angular momentum component $\mathbb{L}_g$ in the complex quaternion space. In other words, a majority of physics quantities are able to be measured in the complex octonion space.

It should be noted that the paper discussed only some simple cases about the contribution of adjoint-fields to the field potential, field strength, field source, angular momentum, torque, and force and so on in the electromagnetic and gravitational field. However it clearly reveals that the application of complex octonion space and octonion operator is capable of exploring availably the field theory about the electromagnetic field, gravitational field, and adjoint-fields. The adjoint-fields of two fundamental fields can be applied to analyze the physics property of dark matter, including the particle category and existence region and so on. In the following study, it is going to probe into the property of E adjoint-field theoretically, and attempt to validate the influence of E adjoint-field source by improving existing experimental devices, and apply the adjoint-fields to the research about the cosmology.

\section*{Acknowledgments}
This project was supported partially by the National Natural Science Foundation of China under grant number 60677039.

\appendix

\section{Double-coordinate system}

The definition of complex coordinates in the complex coordinate system will be seized of one vital revelatory sense, which is of help to restructure the complex octonion coordinate system into the complex quaternion coordinate system with the double-coordinate.

In the complex quaternion coordinate system, for the basis vector $\emph{\textbf{i}}_0$ , the coordinate of physics quantity is $c_0$ ; for the basis vector $\emph{\textbf{i}}_k$ , the coordinate of physics quantity is $c_k$ . And the complex quaternion physics quantity is written as, $( c_0 \emph{\textbf{i}}_0 + \Sigma c_k \emph{\textbf{i}}_k )$. Meanwhile in the complex $S$-quaternion coordinate system, for the basis vector $\emph{\textbf{i}}_0$, the coordinate of physics quantity is $( d_0 \emph{\textbf{I}}_0 )$; for the basis vector $\emph{\textbf{i}}_k$ , the coordinate of physics quantity is $( d_k \emph{\textbf{I}}_0^* )$. And the complex $S$-quaternion physics quantity is written as, $\{ ( d_0 \emph{\textbf{I}}_0 ) \circ \emph{\textbf{i}}_0 + \Sigma ( d_k \emph{\textbf{I}}_0^* ) \circ \emph{\textbf{i}}_k \}$. Herein $c_j$ and $d_j$ are complex numbers.

In the complex quaternion coordinate system with the double-coordinate, it is able to restructure the physics quantity in the complex octonion coordinate system. For the basis vector $\emph{\textbf{i}}_0$ , the double-coordinate of physics quantity is, $( c_0 + d_0 \emph{\textbf{I}}_0 )$; for the basis vector $\emph{\textbf{i}}_k$ , the double-coordinate of physics quantity is, $( c_k + d_k \emph{\textbf{I}}_0^* )$. Eventually in the complex quaternion coordinate system with the double-coordinate, the complex octonion physics quantity can be rewritten as, $\{ ( c_0 + d_0 \emph{\textbf{I}}_0 ) \circ \emph{\textbf{i}}_0 + \Sigma ( c_k + d_k \emph{\textbf{I}}_0^* ) \circ \emph{\textbf{i}}_k \}$. This restructuring will be of great benefit to comprehend and calculate the complex octonion physics quantity.

\section{Measurability}

The complex octonion space is able to be separated into two orthogonal subspaces, the complex quaternion space and $S$-quaternion space. The physics quantity of gravitational field is situated on the complex quaternion space, while the physics quantity of electromagnetic field remains in the complex $S$-quaternion space. According to the octonion multiplication, the product of two complex $S$-quaternion physics quantities belongs to the complex quaternion space, rather than the complex $S$-quaternion space.

For instance, the vector product, of the electric current and velocity in the complex $S$-quaternion space, is one force term (that is, Lorentz force), and belongs to the complex quaternion space. This force term can combine with other force terms in the complex quaternion space, exerting a force on the ordinary matter together. Making use of measuring the applied force of ordinary matters, it is able to figure out the magnitude of the Lorentz force, and therefore work out the magnitude of electric current and of velocity respectively.

In general, the product (in the E adjoint-field) of two complex $S$-quaternion physics quantities will stay at the complex quaternion space, and combine with the relevant complex quaternion physics quantity, making an influence on ordinary matters. In terms of the product characteristics of complex $S$-quaternion physics quantity, it is capable of calculating and measuring some physics quantities of electromagnetic fields, including the electric field intensity, magnetic flux density, Lorentz force, Coulomb force, electric charge and current, angular momentum, scalar-like magnetic moment, radius vector $\mathbb{R}_e$ , velocity $\mathbb{V}_e$ , and the torque and energy between the electric/magnetic dipole moment with the electromagnetic strength and so on (Table 7).

In the complex octonion space, for the fundamental field, a majority of physics quantities are able to be measured already. The physics quantity in the complex quaternion space can be directly measured, while that in the complex $S$-quaternion space may be indirectly surveyed. In particular in terms of measuring and comparing the angular momentum, one can differentiate and identify the contribution of the radius vector and velocity in the complex $S$-quaternion space.

In a similar way, a part of adjoint-field's physics quantities situate in the complex quaternion space, while the rest remains in the complex $S$-quaternion space. Also the most of adjoint-field's physics quantities can be measured. In other words, it is capable of reckoning some physics quantities in the adjoint-field, by means of surveying the contribution of adjoint-field's physics quantities on the field strength, field source, angular momentum, and force and so on.

\begin{table}[h]
\tbl{The measurability of physics quantities relevant to the fundamental field and adjoint-field in the complex octonion space.}
{\begin{tabular}{@{}lll@{}}
\hline
physics~quantity	&  gravitational field     &  electromagnetic field        \\
                    &  (direct measure)        &  (indirect measure)           \\
\hline
field~potential	    &  (not yet)               &  (not yet)                    \\
field~strength	    &  Yes	                   &  Yes                          \\
field~source        &  Yes                     &  Yes                          \\
angular~momentum    &  Yes                     &  Yes                          \\
torque and work     &  Yes                     &  Yes                          \\
force and power     &  Yes                     &  Yes                          \\
radius vector       &  Yes                     &  should be                    \\
velocity            &  Yes                     &  should be                    \\
\hline
\end{tabular}}
\end{table}

\section{Field equations}

\subsection{Gravitational field equations}

Acting the quaternion operator, $\square_g = i \partial_{r0} + \Sigma \emph{\textbf{i}}_k \partial_{rk}$ , on the octonion field strength component, $\mathbb{F}_g = i \textbf{g} / v_0 + \textbf{b}$ , may produce the quaternion equation of gravitational field source, $\mu_g^g \mathbb{S}_g^g = - \square_g^* \circ \mathbb{F}_g$ .

Respectively comparing the scalar and vector parts at both sides of equal sign, in the quaternion equation of gravitational field source, will yield the quaternion gravitational field equations,
\begin{eqnarray}
& \nabla_r^* \cdot \textbf{b} = 0 ~,
\\
& \partial_{r0} \textbf{b} + \nabla_r^* \times \textbf{g} / v_0 = 0 ~,
\\
& \nabla_r^* \cdot \textbf{g} / v_0 = - \mu_g^g S_{g0}^g ~,
\\
& \nabla_r^* \times \textbf{b} - \partial_{r0} \textbf{g} / v_0 = - \mu_g^g \textbf{S}_g^g ~,
\end{eqnarray}
where $\mathbb{S}_g^g = i S_{g0}^g + \textbf{S}_g^g$ . $\textbf{S}_g^g = \Sigma \emph{\textbf{i}}_k S_{gk}^g$ .

Since the gravitational constant $G$ is tiny, and the velocity ratio $(\textbf{v} / v_0)$ is small, the contribution of the linear momentum $\textbf{S}_g^g$ to the gravity can be neglected in general. When the field strength component $\textbf{b}$ and vector potential $\textbf{A}_g$ both are zero in the gravitational field, Eq. (C.3) can be degenerated into the Newton's law of universal gravitation in the classical gravitational theory. Herein $\textbf{v} = \Sigma \emph{\textbf{i}}_k V_{gk}^g$ . $\textbf{A}_g = \Sigma \emph{\textbf{i}}_k A_g^k$ .

\subsection{Electromagnetic field equations}

Acting the quaternion operator, $\square_g = i \partial_{r0} + \Sigma \emph{\textbf{i}}_k \partial_{rk}$ , on the octonion field strength component, $\mathbb{F}_e = i \textbf{E} / v_0 + \textbf{B}$ , can deduce the $S$-quaternion equation of electromagnetic field source, $\mu_e^g \mathbb{S}_e^g = - \square_g^* \circ \mathbb{F}_e$ .

Comparing respectively the scalar and vector parts at both sides of equal sign, in the quaternion equation of electromagnetic field source, will infer the quaternion electromagnetic field equations as follows,
\begin{eqnarray}
& \nabla_r^* \cdot \textbf{B} = 0 ~,
\\
& \partial_{r0} \textbf{B} + \nabla_r^* \times \textbf{E} / v_0 = 0 ~,
\\
& \nabla_r^* \cdot \textbf{E} / v_0 = - \mu_e^g \textbf{S}_{e0}^g ~,
\\
& \nabla_r^* \times \textbf{B} - \partial_{r0} \textbf{E} / v_0 = - \mu_g^g \textbf{S}_e^g ~,
\end{eqnarray}
where $\mathbb{S}_e^g = i \textbf{S}_{e0}^g + \textbf{S}_e^g$ . $\textbf{S}_{e0}^g = \emph{\textbf{I}}_0 S_{e0}^g$ . $\textbf{S}_e^g = \Sigma \emph{\textbf{I}}_k S_{ek}^g$ .

By equivalently transforming those field equations into that in the three-dimensional space, it reveals that the above is identical with the Maxwell's equations in the classical electromagnetic theory.

\subsection{G adjoint-field equations}

Acting the $S$-quaternion operator, $\square_e = i \emph{\textbf{I}}_0 \partial_{R0} + \Sigma \emph{\textbf{I}}_k \partial_{Rk}$ , on the octonion field strength component, $\mathbb{F}_g = i \textbf{g} / v_0 + \textbf{b}$ , can generate the $S$-quaternion equation of G adjoint-field source, $\mu_g^e \mathbb{S}_g^e = - \square_e^* \circ \mathbb{F}_g$ .

In the $S$-quaternion equation of G adjoint-field source, comparing the scalar and vector parts at both sides of equal sign respectively, will conclude the $S$-quaternion G adjoint-field equations,
\begin{eqnarray}
& \nabla_R^* \cdot \textbf{b} = 0 ~,
\\
& \emph{\textbf{I}}_0^* \partial_{R0} \circ \textbf{b} + \nabla_R^* \times \textbf{g} / v_0 = 0 ~,
\\
& \nabla_R^* \cdot \textbf{g} / v_0 = - \mu_g^e \textbf{S}_{g0}^e ~,
\\
& \nabla_R^* \times \textbf{b} - \emph{\textbf{I}}_0^* \partial_{R0} \circ \textbf{g} / v_0 = - \mu_g^e \textbf{S}_g^e ~,
\end{eqnarray}
where $\mathbb{S}_g^e = i \textbf{S}_{g0}^e + \textbf{S}_g^e$ . $\textbf{S}_{g0}^e = \emph{\textbf{I}}_0 S_{g0}^e$ . $\textbf{S}_g^e = \Sigma \emph{\textbf{I}}_k S_{gk}^e$ .

According to the octonion multiplication, the G adjoint-field source, $\mathbb{S}_g^e$ , remains in the complex $S$-quaternion space, $\mathbb{H}_e$ . And it can be combined with the electromagnetic source, $\mathbb{S}_e^g$ .

\subsection{E adjoint-field equations}

Acting the $S$-quaternion operator, $\square_e = i \emph{\textbf{I}}_0 \partial_{R0} + \Sigma \emph{\textbf{I}}_k \partial_{Rk}$ , on the octonion field strength component, $\mathbb{F}_e = i \textbf{E} / v_0 + \textbf{B}$ , can deduce the quaternion equation of E adjoint-field source, $\mu_e^e \mathbb{S}_e^e = - \square_e^* \circ \mathbb{F}_e$ .

In the quaternion equation of E adjoint-field source, comparing the scalar and vector parts at both sides of equal sign respectively, will infer the quaternion E adjoint-field equations,
\begin{eqnarray}
& \nabla_R^* \cdot \textbf{B} = 0 ~,
\\
& \emph{\textbf{I}}_0^* \partial_{R0} \circ \textbf{B} + \nabla_R^* \times \textbf{E} / v_0 = 0 ~,
\\
& \nabla_R^* \cdot \textbf{E} / v_0 = - \mu_e^e S_{e0}^e ~,
\\
& \nabla_R^* \times \textbf{B} - \emph{\textbf{I}}_0^* \partial_{R0} \circ \textbf{E} / v_0 = - \mu_e^e \textbf{S}_e^e ~,
\end{eqnarray}
where $\mathbb{S}_e^e = i S_{e0}^e + \textbf{S}_e^e$ . $\textbf{S}_e^e = \Sigma \emph{\textbf{i}}_k S_{ek}^e$ .

The E adjoint-field source, $\mathbb{S}_e^e$ , stays in the complex quaternion space, $\mathbb{H}_g$ , according to the octonion multiplication. And it can be combined with the gravitational source, $\mathbb{S}_g^g$ .



\begin{thebibliography}{0}    

\bibitem{kuijken}
      K. Kuijken and G. Gilmore,
      {\it Mon. Not. R. Astron. Soc.\/}
      {\bf 239} (1989) 651.

\bibitem{zwicky}
      F. Zwicky,
      {\it Astrophys. J.\/}
      {\bf 86} (1937) 217.

\bibitem{rubin1}
      V. C. Rubin, W. K. Ford, and N. Thonnard,
      {\it Astrophys. J.\/}
      {\bf 238} (1980) 471.

\bibitem{kusenko}
      A. Kusenko,
      {\it Int. J. Mod. Phys. D\/}
      {\bf 16} (2007) 2325.

\bibitem{bertone}
      G. Bertone, D. Hooper, and J. Silk,
      {\it Phys. Rep.\/}
      {\bf 405} (2005) 279.

\bibitem{gupta}
      V. Gupta, R. Kabir, and A. Mukherjee, $et~al$.,
      {\it Int. J. Mod. Phys. D\/}
      {\bf 24} (2015) 1550068.

\bibitem{dietrich}
      J. P. Dietrich, N. Werner, and D. Clowe, $et~al$.,
      {\it Nature\/}
      {\bf 487} (2012) 202.

\bibitem{rubin2}
      V. C. Rubin and W. K. Ford,
      {\it Astrophys. J.\/}
      {\bf 159} (1970) 379.

\bibitem{faber}
      S. M. Faber and R. E. Jackson,
      {\it Astrophys. J.\/}
      {\bf 204} (1976) 668.

\bibitem{koopmans}
      L. V. E. Koopmans and T. Treu,
      {\it Astrophys. J.\/}
      {\bf 583} (2003) 606.

\bibitem{diego}
      J. M. Diego, E. Martinez, and J. L. Sanz, $et~al$.,
      {\it Mon. Not. R. Astron. Soc.\/}
      {\bf 331} (2002) 556.

\bibitem{dekel}
      A. Dekel, F. Stoehr, and G. A. Mamon, $et~al$.,
      {\it Nature\/}
      {\bf 437} (2005) 707.

\bibitem{kozaczuk}
      J. Kozaczuk and S. Profumo,
      {\it Phys. Rev. D\/}
      {\bf 89} (2014) 095012.

\bibitem{cresst}
      CRESST Collab. (G. Angloher, $et~al$.),
      {\it Eur. Phys. J. C\/}
      {\bf 74} (2014) 3184.

\bibitem{edelweiss}
      EDELWEISS Collab. (B. Schmidt, $et~al$.),
      {\it Astropart. Phys.\/}
      {\bf 44} (2013) 28.

\bibitem{munster}
      A. Munster, M. V. Sivers, and G. Angloher, $et~al$.,
      {\it J. Cosmol. Astropart. Phys.\/}
      {\bf 05} (2014) 018.

\bibitem{ghag}
      C. Ghag, D. Yu. Akimov, and H. M. Araujo, $et~al$.,
      {\it Astropart. Phys.\/}
      {\bf 35} (2011) 76.

\bibitem{aprile}
      E. Aprile, F. Agostini, and M. Alfonsi, $et~al$.,
      {\it J. Instrum.\/}
      {\bf 9} (2014) P11006.

\bibitem{amaudruz}
      P.-A. Amaudruz, M. Batygov, and B. Beltran, $et~al$.,
      {\it Astropart. Phys.\/}
      {\bf 62} (2015) 178.

\bibitem{badertscher}
      A. Badertscher, F. Bay, and N. Bourgeois, $et~al$.,
      {\it J. Instrum.\/}
      {\bf 8} (2013) C09005.

\bibitem{weinberg}
      M. D. Weinberg and L. Blitz,
      {\it Astrophys. J. Lett.\/}
      {\bf 641} (2006) L33.

\bibitem{agnes}
      P. Agnes, T. Alexander, and A. Alton, $et~al$.,
      {\it Phys. Lett. B\/}
      {\bf 743} (2015) 456.

\bibitem{lux}
      LUX Collab. (D. S. Akerib, $et~al$.),
      {\it Phys. Rev. Lett.\/}
      {\bf 112} (2014) 091303.

\bibitem{pandax}
      PandaX Collab. (M.-J. Xiao, $et~al$.),
      {\it Sci. China: Phys. Mech. Astron.\/}
      {\bf 57} (2014) 2024.

\bibitem{felizardo}
      M. Felizardo, T. A. Girard, and T. Morlat, $et~al$.,
      {\it Phys. Rev. D\/}
      {\bf 89} (2014) 072013.

\bibitem{archambault}
      S. Archambault, E. Behnke, and P. Bhattacharjee, $et~al$.,
      {\it Phys. Lett. B\/}
      {\bf 711} (2012) 153.

\bibitem{adriani}
      O. Adriani, G. C. Barbarino, and G. A. Bazilevskaya, $et~al$.,
      {\it J. Geophys. Res.: Space Phys.\/}
      {\bf 120} (2015) 3728.

\bibitem{ams}
      AMS Collab. (M. Aguilar, $et~al$.),
      {\it Phys. Rev. Lett.\/}
      {\bf 110} (2013) 141102.

\bibitem{angus}
      G. W. Angus, A. Diaferio, and B. Famaey, $et~al$.,
      {\it Mon. Not. R. Astron. Soc.\/}
      {\bf 436} (2013) 202.

\bibitem{sahu}
      S. Sahu, K. Lochan, and D. Narasimha,
      {\it Phys. Rev. D\/}
      {\bf 91} (2015) 063001.

\bibitem{safarzadeh}
      F. Safarzadeh-Maleki and D. Kamani,
      {\it Phys. Rev. D\/}
      {\bf 90} (2014) 107902.

\bibitem{rama}
      S. K. Rama,
      {\it Phys. Rev. D\/}
      {\bf 89} (2014) 084019.

\bibitem{dellieu}
      L. Dellieu, O. Deparis, and J. Muller, $et~al$.,
      {\it Phys. Rev. Lett.\/}
      {\bf 114} (2015) 024501.

\bibitem{derevianko}
      A. Derevianko	and M. Pospelov,
      {\it Nature: Phys.\/}
      {\bf 10} (2014) 933.

\bibitem{landsberg}
      G. Landsberg,
      {\it Mod. Phys. Lett. A\/}
      {\bf 30} (2015) 1540017.

\bibitem{mcculloch}
      M. E. McCulloch,
      {\it Astrophys. Space Sci.\/}
      {\bf 342} (2012) 575.

\bibitem{ekli}
      E.-K. Li, Y. Zhang, and J.-L. Geng, $et~al$.,
      {\it Astrophys. Space Sci.\/}
      {\bf 355} (2015) 187.

\bibitem{stadnik}
      Y. V. Stadnik and V. V. Flambaum,
      {\it Eur. Phys. J. C\/}
      {\bf 75} (2015) 110.

\bibitem{klimchitskaya}
      G. L. Klimchitskaya and V. M. Mostepanenko,
      {\it Gravit. Cosmol.\/}
      {\bf 21} (2015) 1.

\bibitem{yhli}
      Y.-H. Li, J.-F. Zhang, and X. Zhang,
      {\it Phys. Lett. B\/}
      {\bf 744} (2015) 213.

\bibitem{weng1}
      Z.-H. Weng,
      {\it AIP Adv.\/}
      {\bf 4} (2014) 087103
      [Erratum: {\it ibid.} {\bf 5} (2015) 109901 ].

\bibitem{anastassiu}
      H. T. Anastassiu, P. E. Atlamazoglou, and D. I. Kaklamani,
      {\it IEEE T. Antenn. Propag.\/}
      {\bf 51} (2003) 2130.

\bibitem{morita}
      K. Morita,
      {\it Prog. Theor. Phys.\/}
      {\bf 117} (2007) 501.

\bibitem{doria}
      F. A. Doria,
      {\it Lett. Nuovo Cimento\/}
      {\bf 14} (1975) 480.

\bibitem{majernik1}
      V. Majernik,
      {\it Adv. Appl. Clifford Al.\/}
      {\bf 9} (1999) 119.

\bibitem{gogberashvili}
      M. Gogberashvili,
      {\it J. Phys. A\/}
      {\bf 39} (2006) 7099.

\bibitem{mironov}
      V. L. Mironov and S. V. Mironov,
      {\it J. Math. Phys.\/}
      {\bf 50} (2009) 012901.

\bibitem{demir}
      S. Demir, M. Tanisli, and T. Tolan,
      {\it Int. J. Mod. Phys. A\/}
      {\bf 28} (2013) 1350112.

\bibitem{brumby1}
      S. P. Brumby, B. E. Hanlon, and G. C. Joshi,
      {\it Phys. Lett. B\/}
      {\bf 401} (1997) 247.

\bibitem{brumby2}
      S. P. Brumby and G. C. Joshi,
      {\it Found. Phys.\/}
      {\bf 26} (1996) 1591.

\bibitem{majernik2}
      V. Majernik,
      {\it Gen. Rel. Grav.\/}
      {\bf 36} (2004) 2139.	

\bibitem{urui1}
      S. Furui,
      {\it Int. J. Mod. Phys. A\/}
      {\bf 27} (2012) 1250158.	

\bibitem{urui2}
      S. Furui,
      {\it Few-Body Syst.\/}
      {\bf 54} (2013) 2097.	

\bibitem{urui3}
      S. Furui,
      {\it Few-Body Syst.\/}
      {\bf 55} (2014) 1083.	

\bibitem{chanyal}
      B. C. Chanyal, V. K. Sharma, and O. P. S. Negi,
      {\it Int. J. Theor. Phys.\/}
      {\bf 54} (2015) 3516.	

\bibitem{weng2}
      Z.-H. Weng,
      {\it Int. J. Mod. Phys. D\/}
      {\bf 24} (2015) 1550072.	

\end{thebibliography}
\end{document}